\documentclass[12pt]{article}
\usepackage{amssymb}
\usepackage{amsmath}
\usepackage{verbatim} 
\usepackage{graphicx,epsfig}
\usepackage{epsfig}

\newcommand{\be}{\begin{equation}}
\newcommand{\ee}{\end{equation}}
\newcommand{\bea}{\begin{eqnarray}}
\newcommand{\eea}{\end{eqnarray}}
\newcommand{\beas}{\begin{eqnarray*}}
\newcommand{\eeas}{\end{eqnarray*}}
\newcommand{\nn}{\nonumber}

\newcommand{\half}{{1\over 2}}

\begin{document}
\numberwithin{equation}{section}

\baselineskip 14 pt
\parskip 12 pt 

\begin{titlepage}

\begin{center}

\vspace{5mm}

{\Large \bf Boundary Terms, Variational Principles and Higher Derivative Modified Gravity }

\vspace{5mm}

Ethan Dyer\footnote{esd2107@columbia.edu} and Kurt Hinterbichler\footnote{kurth@phys.columbia.edu}

\vspace{2mm}

{\small \sl Institute for Strings, Cosmology and Astroparticle Physics} \\
{\small \sl and Department of Physics} \\
{\small \sl Columbia University, New York, NY 10027 USA}

\end{center}

\vskip 1.0 cm

\noindent
We discuss the criteria that must be satisfied by a well-posed variational principle.  We clarify the role of Gibbons-Hawking-York type boundary terms in the actions of higher derivative models of gravity, such as $F(R)$ gravity, and argue that the correct boundary terms are the naive ones obtained though the correspondence with scalar-tensor theory, despite the fact that variations of normal derivatives of the metric must be fixed on the boundary.  We show in the case of $F(R)$ gravity that these boundary terms reproduce the correct ADM energy in the hamiltonian formalism, and the correct entropy for black holes in the semi-classical approximation.  

\end{titlepage}
\setcounter{footnote}{0}

\section{Introduction and outline\label{Intro}}

The Einstein-Hilbert action for general relativity (GR) is
\be S_{EH}\sim \int d^4 x \ \sqrt{-g}R.\ee
When varying this action, one finds surface contributions that must vanish if the action is to be stationary.  The surface contributions contain the metric variation $\delta g_{\mu\nu}$ and variations of the derivatives of the metric $\delta( \partial_\sigma g_{\mu\nu})$.  Setting $\delta g_{\mu\nu}=0$ on the boundary is not sufficient to kill all the surface contributions.  Fixing both the metric and the derivatives of the metric on the boundary is uncomfortable, however it may not be initially obvious why this is so, or that there exists a unique and correct prescription for dealing with it.

Gibbons, Hawking, and York proposed adding the trace of the extrinsic curvature of the boundary, $K$, to the action \cite{Gibbons:1976ue,York:1972sj}. With this modification the action takes the form
\be S=S_{EH}+S_{GHY}\sim \int d^4 x \ \sqrt{-g}R+2\oint d^{3}x\sqrt{|h|}K.\ee
$h$ is the determinant of the induced metric on the boundary.  This modification is appealing because the variation of the Gibbons-Hawking-York (GHY) boundary term cancels the terms involving $\delta( \partial_\sigma g_{\mu\nu})$, and so setting $\delta g_{\mu\nu}=0$ becomes sufficient to make the action stationary. This has been widely accepted as the correct modification to the action.

This modification raises some questions.  Is the modification unique? Is it necessarily incorrect to require  $\delta (\partial_\sigma g_{\mu\nu})=0$?  If this is incorrect, why is it appropriate to require the fixing of all components of $g_{\mu\nu}$, when the graviton has only two degrees of freedom?  More generally,  what criteria determine which quantities to fix on the boundary?  Should it be related to the number of degrees of freedom in the theory?

Despite these questions the GHY term is desirable, as it possesses a number of other key features. The term is required to ensure the path integral has the correct composition properties   \cite{Gibbons:1976ue}. When passing to the hamiltonian formalism, it is necessary to include the GHY term in order to reproduce the correct Arnowitt-Deser-Misner (ADM) energy \cite{Hawking:1995fd}.  When calculating black hole entropy using the euclidean semiclassical approach, the entire contribution comes from the GHY term \cite{Brown:1992bq}. These considerations underscore the necessity of a complete understanding of boundary terms.

Higher derivative theories of gravity have attracted attention recently, mostly as modifications to GR that have the potential to explain cosmic acceleration without vacuum energy.  A simple example of such a modification is $F(R)$ gravity, where the action becomes some arbitrary function, $F$, of the Ricci scalar (for reviews, see \cite{Sotiriou:2008rp,Nojiri:2006ri}), 
\be S\sim \int d^4 x \ \sqrt{-g}F(R).\ee
It is well known that this theory is dynamically equivalent to a scalar-tensor theory \cite{Teyssandier:1983zz,Whitt:1984pd,Barrow:1988xh,Barrow:1988xi,Chiba:2003ir} .  By extending this equivalence to the GHY term, we find that the $F(R)$ action is left with a boundary term, 
\be \int d^4 x \ \sqrt{|g|}F(R)+2\oint d^{3}x\sqrt{|h|}F'(R)K.\ee
This boundary term must be present if the correspondence to scalar-tensor theory is to hold at the boundary.  

This term has been arrived at before, both directly and indirectly, and in several different contexts \cite{Balcerzak:2008bg,Barth:1984jb,Casadio:2001ff,Madsen:1989rz,Nojiri:1999nd}.  There has been some confusion, as this boundary term does not allow $\delta (\partial_\sigma g_{\mu\nu})$ to remain arbitrary on the boundary.  Various ways around this have been attempted, for example, when restricting to maximally symmetric backgrounds a boundary term for $F(R)$ theory can be found that allows $\delta (\partial_\sigma g_{\mu\nu})$ to remain arbitrary on the boundary \cite{Madsen:1989rz}.  

$F(R)$ theory is a higher derivative theory, however, and the derivatives of the metric encode true degrees of freedom.  $\delta (\partial_\sigma g_{\mu\nu})$ should not remain arbitrary on the boundary.  Instead, $\delta (\partial_\sigma g_{\mu\nu})$ must be subject to the constraint that the variation of the four-dimensional Ricci scalar be held fixed on the boundary.  This corresponds to holding the scalar field fixed in the equivalent scalar-tensor theory.   We will show that the above boundary term reproduces the expected ADM energy upon passing to the hamiltonian formalism, and the expected entropy for black holes.  

In what follows, we will elaborate upon the above in detail.

An outline of this paper is as follows.  In section \ref{variational}, we go over the criteria that are necessary to have a well-posed variational principle.  In section \ref{examples}, we examine several toy examples of lagrangians in classical mechanics that share some of the features of the more complicated GR case.  We discuss what adding total derivatives can do to the variational principle, and what happens in higher derivative theories.  We also discuss the complications due to constraints and gauge invariance, and we work out the case of electromagnetism.  In section \ref{gibbonshawking}, we review the GHY term, its variation, and how it renders the GR action well-posed.  In section \ref{scalartensor}, we discuss higher derivative modified gravity theories, scalar-tensor theories, and their equivalence, using $F(R)$ theory as the prime example.  We find the boundary terms for $F(R)$ theory using this equivalence.  In section \ref{hamiltonian}, we derive the hamiltonian formulation of scalar-tensor theory and $F(R)$ theory, keeping all boundary terms, and show how the boundary term is essential for obtaining the ADM energy.  In section \ref{blackholes}, we calculate the entropy of a Schwartzschild black hole in $F(R)$ theory, using the euclidean semi-classical approach, and compare the result to the Wald entropy formula.  We conclude in section \ref{conclusion}, and some formulae, theorems and technicalities are relegated to the appendices.  

\textbf{Conventions:}  We use the conventions of Carroll \cite{Carroll:2004st}.   These include using the mostly plus metric signature $(-,+,+,+,\ldots)$, and the following definitions of the Riemann and Ricci tensors, ${R^\kappa} _{\lambda \mu \nu} = \partial _\mu \Gamma^\kappa _{\nu 
\lambda} - \partial_\nu \Gamma ^\kappa _{\mu \lambda} + \Gamma ^\kappa _{\mu \eta} \Gamma^\eta _{\nu 
\lambda} - \Gamma 
^\kappa _{\nu \eta}\Gamma ^\eta _{\mu \lambda} $, $R_{\lambda\nu}={R^\mu} _{\lambda \mu \nu}$.  The weight for anti-symmetrization and symmetrization is, e.g. $A_{[\mu\nu]}=\half\left(A_{\mu\nu}-A_{\nu\mu}\right)$.  For spacelike hypersurfaces, the normal vector will always be inward pointing.  For timelike hypersurfaces, it will be outward pointing.  The dimension of spacetime is $n$.  Further conventions and notation for foliations of spacetime are laid out in appendices \ref{31appendix} and \ref{foliationappendix}.

\section{\label{variational} What makes a variational principle well-posed?}

A traditional approach to field theory is to integrate by parts at will, ignoring boundary contributions.  The reasoning is that if there are no physical boundaries in the space under consideration, or if they are so far away that their effects can be expected not to interfere with the system under study, then they can be ignored.  

If one is interested only in the local equations of motion of a theory, this is a valid approach.  In this case, one uses the action only as a formal device for arriving at the equations of motion.   Denoting the fields collectively by $\phi^i$, one writes down the lagrangian, which is a local density,
\be {\cal L}\left([\phi],x\right).\ee
Here $[\phi]$ stands for dependence on $\phi^i$ and any of its higher derivatives, $[\phi]\equiv \phi^i,\partial_\mu\phi^i,\partial_\mu\partial_\nu\phi^i,\ldots$, and $x$ are the coordinates on spacetime.  One then defines the equations of motion to be the Euler-Lagrange equations 
\be \label{eulerlagrange} {\delta ^{EL}{\cal L}\over \delta\phi^i}=0,\ee
where
\[ \frac{\delta^{EL}}{\delta\phi^i}=\frac{\partial}{\partial\phi^i}-\partial_\mu\frac{\partial}{\partial(\partial_\mu\phi^i)}+\partial_\mu\partial_\nu\frac{\partial^S}{\partial(\partial_\mu\partial_\nu\phi^i)}-\cdots\]
is the Euler-Lagrange derivative and the symmetric derivative is defined by 
\be \frac{\partial^S}{\partial(\partial_{\mu_1}\ldots\partial_{\mu_k}\phi^i)}\partial_{\nu_1}\ldots\partial_{\nu_k}\phi^j=\delta^i_j\delta^{\mu_1}_{(\nu_1}\cdots \delta^{\mu_k}_{\nu_k)}.\ee
(The symmetric derivative is to avoid over-counting multiple derivatives which are not independent.) 

All the relevant local theory, i.e. equations of motion, symmetries, conserved currents, gauge symmetries, Noether identities, etc., can be developed in terms of the lagrangian density alone, without ever writing an integral and without consideration of boundary contributions.   For example, Noether's theorem can be stated as: an infinitesimal transformation $\delta\phi^i\left([\phi],x\right)$ is a symmetry if the lagrangian changes by a total derivative, 
\be \delta\mathcal{L}= \partial_\mu k^\mu,\ee
 from which one can show directly that the following current is conserved on shell
\be j^\mu=-k^\mu+\sum_{r=1}^n\partial_{\mu_1}\ldots\partial_{\mu_{r-1}}\delta\phi^i\sum_{l=r}^n(-1)^{l-r}\partial_{\mu_r}\ldots\partial_{\mu_{l-1}}\frac{\partial^S\mathcal{L}}{\partial(\partial_\mu\partial_{\mu_1}\ldots\partial_{\mu_{l-1}}\phi^i)}.\ee
(This reduces to $j^\mu=\delta\phi^i\frac{\partial\mathcal{L}}{\partial(\partial_\mu\phi^i)}-k^\mu$ in the usual case where the lagrangian depends at most on first derivatives of the fields.)\footnote{If one wishes, still keeping with this line of thought, the action integral can be introduced as a formal device, 
\be \label{standardform} S=\int  d^nx\ {\cal L}\left([\phi],x\right).\ee
The integration region need not be specified, and the equations of motion are obtained by setting $\delta S=0$ and integrating by parts ignoring all boundary contributions.}

Without specifying boundary conditions, the Euler-Lagrange equations (\ref{eulerlagrange}) typically have many solutions (if there are no solutions, the lagrangian is said to be inconsistent).  To select out the true solution, boundary conditions must be set.  There should be some class of boundary conditions that render the system well-posed.  A class of boundary conditions is well-posed if, given any choice of boundary conditions from this class, there exists a unique solution to the system compatible with that choice.  The equations of motion are typically hyperbolic, so the class will generally involve all possible choices of initial conditions and velocities for the fields, and all possible choices of spatial boundary conditions at all times.  There may be several different classes of well-posed data, but each class will involve specifying the same number of boundary data.  This number is the number of degrees of freedom in the theory.  

The choice of spatial boundary conditions generally corresponds to a choice of ``vacuum state'' for the theory.  For example, in GR, if spatial boundary conditions (fall-off behavior for the fields, in this case) are chosen so that the spacetime is asymptotically flat, we find ourselves in the asymptotically flat vacuum of the theory.  A choice of initial condition generally corresponds to a choice of state within the vacuum.  For example, both Minkowski space and the Schwartzschild black hole have the same spatial boundary behavior, so they are both in the asymptotically flat vacuum of GR, but they have different initial data, and so they represent different states.  Thus, from the point of view that the action is just a formal device to arrive at the local equations of motion, information about the possible vacua of the theory, and the space of states in each vacuum, is not encoded directly in the action, only indirectly through the boundary conditions required of the equations of motion.   

\begin{figure}[h!]
\begin{center}
\includegraphics[height=2.5in]{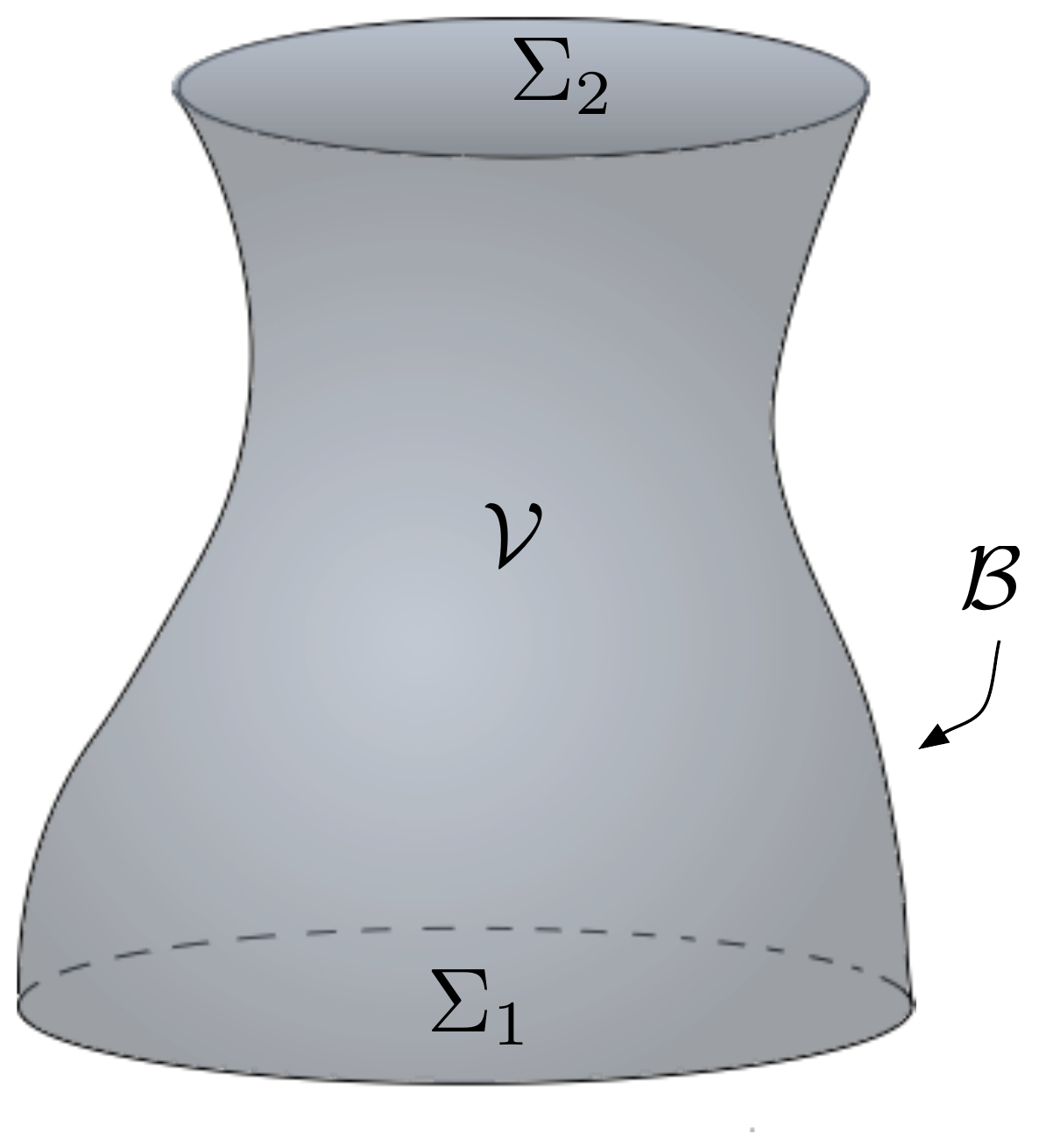}
\caption{Region over which a field theory is defined.}
\label{tube}
\end{center}
\end{figure}

Ideally, the action should do more.  We want to say that the true field configuration, among all field configurations with given boundary data, extremizes the action.  Consider a spacetime region, ${\cal V}$, such as that in figure \ref{tube}.  The boundary of this region consists of a timelike part ${\cal B}$, (which may be at infinity), which we will refer to as the \emph{spatial boundary}, and two spacelike edges, $\Sigma_1$ and $\Sigma_2$, which we will refer to as the \emph{endpoints}.   We consider configurations of the fields $\phi^i$ in ${\cal V}$ that have some given boundary values, $\phi^i({\cal B})$, as well as initial and final values, $\phi^i(\Sigma_{1})$, $\phi^i(\Sigma_2)$ (possibly involving derivatives of the fields or other combinations), chosen from a class which is well-posed, meaning that for each choice in the class there exists a unique solution to the equations of motion.\footnote{Well-posedness of hyperbolic equations usually requires \emph{initial} data, such as the values of $\phi^i$ and its time derivatives at $\Sigma_{1}$, rather than endpoint data such as the value of $\phi^i$ at $\Sigma_{1}$ and $\Sigma_{2}$.  The action naturally wants endpoint data, however.  As we will elaborate on in the next section, endpoint data is usually just as well posed as initial data, with the exception of ``unfortunate choices'' which are measure zero in the space of all choices.}

The action is a functional of field configurations in ${\cal V}$,
\be S=S[\phi({\cal V})],\ee
in the form of an integral over over ${\cal V}$ of the lagrangian ${\cal L}\left([\phi],x\right),$
\be \label{standardform} S[\phi({\cal V})]=\int _{\cal V}d^nx\ {\cal L}\left([\phi],x\right).\ee
The criteria for a well-posed action should be the following.  \emph{If we restrict the action to those field configurations consistent with our choice of boundary and initial data,} 
\be S=\left.\int _{\cal V} d^nx\ {\cal L}\left([\phi],x\right)\right|_{\phi^i\ \text{consistent with data}\ \phi^i({\cal B}),\phi^i(\Sigma_{1}),\phi^i(\Sigma_{2})},\ee 
\emph{the unique solution to the Euler-Lagrange equations should be the only extremum of the action.}  It is important that the space of paths over which we vary be \emph{all} paths that have the given boundary data.

Of course, a necessary condition for the action to be stationary is that the field satisfy the Euler-Lagrange equations 
\be {\delta ^{EL}{\cal L}\over \delta\phi^i}=0.\ee
For the action to be truly stationary, any boundary contributions arising from the variation must vanish. No conditions other than those implicit in the class of well-posed data may be used in checking whether boundary contributions vanish.    

There will be many possible choices of the boundary data $\phi^i({\cal B})$ (corresponding to different choices of vacuum) and endpoint data $\phi^i(\Sigma_{1})$, $\phi^i(\Sigma_2)$ (corresponding to different choices of the state) within a well-posed class.  Each such choice leads to a different set of paths to consider in the action, and a well-posed action will give a unique stationary field configuration for each. \emph{The action, when evaluated on the classical path $\phi^i_c$, is thus a well defined functional on the class of possible boundary data},
\be \left.S[\phi({\cal B}),\phi(\Sigma_{1}),\phi(\Sigma_{2})]\right|_{\phi=\phi_c}=\left.\int _{\cal V}d^nx\ {\cal L}\left([\phi],x\right) \right|_{\phi=\phi_c}.\ee
The number of physical degrees of freedom in the theory is the number of free choices of $\phi^i(\partial{\cal V})$ (modulo gauge invariance, see later), i.e. the size of the class of well-posed boundary data.  

These criteria are essential to setting up a Hamilton-Jacobi theory, where the stationary value of the action, as a function of endpoint data, gives the canonical transformation that solves the equations of motion \cite{Goldstein}.  These criteria are also essential to setting up a quantum theory based on the path integral. \cite{Gibbons:1976ue}.

We will now turn to some examples.  We will look at some actions which are well-posed in the sense we have just described, and some which are not.  We will see that adding boundary terms or, equivalently, total derivatives to the action affects whether the action can be well-posed.  Second class constraints and gauge freedom complicate the issue, and we will have more to say about these as well.  

\section{\label{examples} Some Toy Examples}
We start with some simple examples from point particle mechanics.  These are one-dimensional field theories, with fields $q^i(t)$ that depend on one parameter, the time $t$.  The issue of spatial boundary data does not arise in this case; it is only the endpoint data, $q^i(t_1)$ and $q^i(t_2)$ that we need to worry about.\footnote{ In this sense, one dimensional theories do not possess different vacua, only a space of states, given by the various values the endpoint data are allowed to take.}  The number of degrees of freedom in the theory is half the number of free choices that may be made in specifying endpoint data.  

\subsection{Standard classical mechanics}

The best example of a well-defined variational principle is an ordinary unconstrained lagrangian system for $n$ variables $q^i(t)$, $i=1,\ldots, n$ without higher derivatives,
\be S[q^i_1,q^i_2,t_1,t_2]=\int _{t_1}^{t_2}dt\ L(q^i,\dot q^i,t),\ee
\be\label{invertible} \det \left({\partial^2 L\over \partial \dot q^i\partial \dot q^j}\right)\not=0.\ee
The action is a function of the endpoint data $q^i_1,q^i_2$, and the endpoint times $t_1,t_2$.  The set of paths over which we vary is all $q^i(t)$ satisfying $q^i(t_1)=q^i_1$ and $q^i(t_2)=q^i_2$. Varying the action we have 
\be \delta S=\int _{t_1}^{t_2}dt\ \left[{\partial L\over \partial q^i}-{d\over dt}{\partial L\over \partial \dot q^i}\right]\delta q^i+\left.{\partial L\over \partial \dot q^i}\delta q^i \right|_{t_1}^{t_2}.\ee

The boundary variation vanishes since the $q^i$ are fixed there, so the necessary and sufficient condition for the action to be stationary is that the Euler-Lagrange equations of motion hold, 
\be {\partial L\over \partial q^i}-{d\over dt}{\partial L\over \partial \dot q^i}=0,\ee
or, upon expanding out the time derivative,
\be \label{secondorder} {\partial^2 L\over \partial \dot q^i\partial \dot q^j}\ddot q^j={\partial L\over \partial q^i}- {\partial^2 L\over \partial \dot q^i\partial q^j}\dot q^j.\ee
Because of (\ref{invertible}), we can solve algebraically for the highest derivative $\ddot q^i$ in terms of the lower derivatives, and we have a well formed initial value problem:  given a choice from the class of data $q^i(t_1)$ and $\dot q^i(t_1)$, we will have a unique solution to the equations of motion.  

However, the action principle is not well suited to the initial value class.  It is suited to the class where we choose endpoint data, i.e. where we fix $q^i$ on the two endpoints but do not specify the $\dot q^i$ anywhere.  For a second order differential equation such as (\ref{secondorder}), the endpoint problem will be well posed for all but a few ``unfortunate'' choices of the endpoint data (borrowing the language of \cite{Henneaux:1992ig}).  An example of such an unfortunate choice is the following.  Take the action of a simple harmonic oscillator in one dimension, with unit mass and angular frequency $\omega$, $S=\int _{t_1}^{t_2}dt\ \half\left(\dot q^2-\omega^2 q^2\right)$.  The equation of motion is $\ddot q+\omega^2 q=0$.  If the time interval is chosen to be a half integer multiple of the period of the oscillator, $t_2-t_1={n\pi\over \omega}$, $n=1,2,\ldots$, and we choose $q_1=q_2=0$, then there will be an infinite number of distinct solutions to the equations of motion compatible with the endpoint data.  This reflects the fact that the period of oscillations is independent of the amplitude.  It is not hard to think of other such examples of ``unfortunate'' choices.  The problem is that the equations of motion want \emph{initial conditions}, but the action principle wants \emph{endpoint} conditions, and the result is a failure of the action principle to be well defined for all choices.  

These unfortunate choices will make up a set of measure zero among all possible choices for the endpoint data.  This is because finding degeneracies in the initial value formulation is equivalent to an eigenvalue problem over the interval $t_2-t_1$ and the eigenvalues are discrete.  We will still consider such a variational principle well-posed.  With this caveat, the action principle above, using the class of endpoint data, is well-posed, because for almost any choice of endpoint data, $q^i_1,q^i_2,t_1,t_2$, there is a unique solution to the equation of motion, corresponding to the unique path which extremizes the action.  The action, evaluated on this path, becomes a well defined functional of $q^i_1,q^i_2,t_1,t_2$ almost everywhere, and so the number degrees of freedom (i.e. the size of the class of data), or the possibility of differentiating the action with respect to endpoint data is not affected.   There are $2n$ free choices among the field endpoint values, so the model has $n$ degrees of freedom.  

Going to the hamiltonian poses no problem.  We introduce fields $p_i(t)$, 
\be \label{pdef} p_i={\partial L\over \partial \dot q^i},\ee
which by virtue of (\ref{invertible}) can be inverted to solve for $\dot q^i=\dot q^i(p,q).$
We then write the action 
\be S=\int _{t_1}^{t_2}dt\ p_i\dot q^i-H,\ee
where 
\be H(p,q)=\left.\left(p_i\dot q^i-L\right)\right|_{\dot q^i=\dot q^i(p,q)}.\ee
The variation with respect to the $p_i$'s is done without fixing the endpoints, so their endpoint values are not specified, and the action is not a function of them.  The equations of motion for $p$ are (\ref{pdef}), which upon plugging back into the action reproduces the original lagrangian action (the transition to the hamiltonian is a special case of the fundamental theorem of auxiliary variables, see appendix \ref{fundamentaltheorem}).  Note that adding a total derivative, ${d\over dt}F(q)$, of any function of the $q^i$ to the original lagrangian does not change the variational principle or the equations of motion, but does change the canonical momenta in a way that amounts to a canonical transformation.

\subsection{Constraints}

Consider now a lagrangian that does not satisfy $\det \left({\partial^2 L\over \partial \dot q^i\partial \dot q^j}\right)\not=0$.  This is the case for essentially all theories of interest to most high energy physicists, with the exception of scalar field theories on fixed backgrounds.  The theory then has constraints, and the Dirac constraint algorithm or some equivalent method must be applied \cite{Dirac,Henneaux:1992ig}.  

As an example, consider a theory with a free point particle of mass $m$, labeled by $q_1$, along with a harmonic oscillator of mass $M$ and angular frequency $\omega$, labeled by $q_2$, coupled through a derivative interaction $\lambda\dot q_1\dot q_2$,
\be S=\int _{t_1}^{t_2}dt\ \half m \dot q_1^2+\half M\dot q_2^2-\half M\omega^2 q_2^2+\lambda\dot q_1 \dot q_2.\ee
This lagrangian is free of constraints except when $\lambda=\pm \sqrt{Mm}$, in which case there is a single primary constraint and a single secondary constraint, which taken together are second class.   It is easy to see why these values of $\lambda$ are special, because for these values we can factor the action,
\be S=\int _{t_1}^{t_2}dt\ \half \left(\sqrt{m}\dot q_1\pm \sqrt{M}\dot q_2\right)^2-\half M\omega^2 q_2^2,\ee
after which a change of variables $q'\equiv \sqrt{m} q_1\pm \sqrt{M} q_2$ shows that $q'$ behaves as a free point particle while $q_2$, which has no kinetic term, is constrained to be zero.  

Varying the action, we find the endpoint term
\be  \left.\left(\sqrt{m}\dot q_1\pm \sqrt{M}\dot q_2\right)\delta \left(\sqrt{m} q_1\pm \sqrt{M} q_2\right) \right|_{t_1}^{t_2}.\ee
The action is stationary if we choose the quantity $\sqrt{m} q_1\pm \sqrt{M} q_2$ to be fixed on the endpoints.  $q_2$ must be chosen to be zero, so choosing this quantity is equivalent to choosing $q_1$. 

This illustrates what should be the case in general with constrained systems.   The variations of the unconstrained variables are fixed on the boundary, and the constrained variables are not fixed.  Each choice of endpoint data for the unconstrained variables determines a choice of data for the constrained variables, which then determines a unique solution to the equations of motion.  The action is thus a function of the endpoint data for the unconstrained variables.  The action should be such that even though all  variables may appear in the endpoint variation, the number that must be fixed to make the action stationary is fewer, equal in number to the number of unconstrained degrees of freedom.

\subsection{Gauge invariance}

Gauge invariance also complicates the well-posedness of variational principles.  Consider the following action for two variables $q_1$ and $q_2$.  
\be S=\int _{t_1}^{t_2}dt\ \half\left(q_1-\dot q_2\right)^2 .\ee
Varying with respect to $q_1$ gives the equation of motion $q_1=\dot q_2$ with no endpoint contribution.  $q_1$ can thus be eliminated as an auxiliary field, and plugging this back into the action gives identically zero.  The general solution to the equation of motion is to let $q_2$ be a completely arbitrary function of $t$, and then set $q_1=\dot q_2$.  The presence of arbitrary functions in the general solution to the equations of motion is the hallmark of a gauge theory, and indeed this theory is invariant under the gauge transformation $\delta q_1=\dot \epsilon$, $\delta q_2=\epsilon$, where $\epsilon(t)$ is an arbitrary function of $t$.  

Arbitrary functions in the solution also indicate that the action principle is not well defined as it stands. Given endpoint data, we can make a gauge transformation that changes the solution somewhere in the bulk, away from the endpoints, and so the solution for the given endpoint data will not be unique.  We salvage uniqueness by identifying as equivalent all solutions which differ by a gauge transformation.  For example, given any $q_2(t)$,  we can, without changing the endpoint values $q_2(t_1),q_2(t_2)$, bring it to the gauge $\ddot q_2=0$, by making a gauge transformation with the gauge parameter $\epsilon(t)$, obtained by solving $\ddot\epsilon+\ddot q_2=0$ subject to the endpoint conditions $\epsilon(t_1)=\epsilon(t_2)=0$. Then, given any choice of $q_2(t_1),q_2(t_2)$, there is a unique solution compatible with $\ddot q_2=0$.  

The gauge condition $\ddot q_2=0$ still allows residual gauge transformations, those where $\epsilon(t)$ is linear in $t$.  These gauge transformations do not vanish at the endpoints.   If a gauge transformation does not vanish at the endpoints, we must identify endpoint data that differ by such a gauge transformation.  In the case above, $q_2(t_1)$ and $q_2(t_2)$ can be set to any desired value by a residual gauge transformation, so all the $q_2$ endpoint data is to be identified into a single equivalence class.  The $q_1$ data is constrained, so in total this model has only a single state, and carries no degrees of freedom. 

The variational principle is still well-posed.  The action is stationary without fixing the constrained variable $q_1$.  The unconstrained variable $q_2$ must be fixed, so the action is a function of its chosen boundary values.  However, gauge invariance of the action ensures that the action takes the same value over all the gauge identified endpoint data, so it is well defined on the equivalence classes.\footnote{In general, gauge transformations may force the identification of other data besides the endpoint data.  An example is the lagrangian for a relativistic point particle which contains reparametrization invariance of the world-line.  In this case, it is the time interval $t_2-t_1$ that is identified \cite{Teitelboim:1981ua}.  In the case of general relativity, there are gauge transformations of both kind.  Spatial diffeomorphisms(generated by the momentum constraints) generate equivalences among the endpoint data, whereas the diffeomorphisms generated by the hamiltonian constraint generates equivalences among the boundary data. }

\subsection{Field theory/spatial boundaries}

As an example of a field theory with gauge invariance, we will show how the action for electromagnetism is well-posed.  Consider Maxwell's equations for a vector field $A_\mu$, to be solved in a region such as that shown in figure \ref{tube},
\be \square A_\mu-\partial_\mu(\partial_\nu A^\nu)=0.\ee
They are gauge invariant under $A_\mu\rightarrow A_\mu+\partial_\mu\Lambda$ for any function $\Lambda(x)$.  

We wish to know what boundary data are required to make this system well-posed.  Start by fixing Lorentz gauge $\partial_\mu A^\mu=0$ in the bulk, by choosing $\Lambda$ to satisfy $\square \Lambda=-\partial_\mu A^\mu$.  The equations of motion are then equivalent to
\be \square A^\mu=0,\ \ \ \partial_\mu A^\mu=0.\ee
These still allow for a residual gauge transformation $A_\mu\rightarrow A_\mu+\partial_\mu\Lambda$ where $\Lambda$ satisfies $\square \Lambda=0$ in the bulk.  Solving the equation $\square \Lambda=0$ requires specifying $\Lambda$ on the boundary, so the residual gauge transformations will be transformations of the data on the boundary, which will be used to generate equivalence classes.   

The wave equation $\square A^\mu=0$ requires specifying $A_\mu$ on $\Sigma _1$, $\Sigma_2$ and $ {\cal B}$ (except for ``unfortunate'' choices).   Differentiating, we have $\square (\partial_\mu A^\mu)=0$, which is the wave equation for the quantity $\partial_\mu A^\mu$.  Thus, if we fix boundary conditions for $A_\mu$,  then extend into the bulk using $\square A^\mu=0$, we need only check that $\partial_\mu A^\mu=0$ on the boundary, as the extension will then automatically satisfy $\partial_\mu A^\mu=0$ in the bulk.

For a given choice of $A^{\mu}$ on the boundary, $\partial_{\mu}A^{\mu}$ is entirely specified by Laplace's equation $\square A^\mu=0$, thus there will be a single constraint on the allowed choices of $A_\mu$ on the boundary to ensure $\partial_{\mu}A^{\mu}=0$ there.  To see the form of the constraint, note that the perpendicular derivatives of the $A_\mu$ are not set as part of the boundary data, but are determined by extending the solution into the bulk.   For instance, we can write
\be \label{fconstraint} \partial_{\perp}A_{\perp}=f(\left.A_{\perp}\right|_{\partial {\cal V}},\left.A_{\parallel}\right|_{\partial {\cal V}}), \ee
where $f$ is some function of the components of $A^{\mu}$ on the boundary specified by solving Laplace's equation.  Therefore we need only impose the restriction
\be\left[ \partial_{\parallel}A_{\parallel}+f(A_{\perp},A_{\parallel})\right]_{\partial{\cal V}}=0 \ee
on the boundary data.  For every choice of boundary data satisfying this restriction, there is a unique solution to Maxwell's equations in Lorentz gauge.  

We can think of this condition as determining $A_\perp$, given $A_\parallel$, so we obtain a unique solution given any choice of $A_\parallel$, with $A_\perp$ determined through the constraint.  To have a well defined variational problem given this choice of the free data, the action should be stationary with only $A_\parallel$ fixed on the boundary.   

We now return to the residual gauge freedom. A gauge transformation such that $\square\Lambda=0$ is still permissible.  These residual gauge transformations leave the constraint (\ref{fconstraint}) invariant, and serve to identity the unconstrained boundary data into equivalence classes.  We can select a member of each class by fixing the residual gauge.  For example, we can use it to set  $A_{\perp}=0$.  We then have two independent functions that must be specified, namely $A_\parallel$, subject to the single constraint (\ref{fconstraint}) . These two functions correspond to the two polarization states of the photon.

Consider now the Maxwell action in this volume,
\be S=\int_{\cal V}d^4x\ -\frac{1}{4} F_{\mu\nu}F^{\mu\nu},\ \ \ F_{\mu\nu}\equiv\partial_\mu A_\nu-\partial_\nu A_\mu. \ee
Variation  gives
\be \int_{\cal V}d^4x\ \partial_\mu F^{\mu\nu}\delta A_\nu-\oint_{\partial{\cal V}}d^3x\ n_{\mu}F^{\mu\nu}\delta A_\nu.\ee
$n^\mu$ is the outward pointing normal on the timelike surface ${\cal B}$, and the inward pointing normal on the spacelike surfaces $\Sigma_1$ and $\Sigma_2$.  Notice that the action is stationary when only $A_\parallel$ is held fixed at the boundaries, due to the anti-symmetry of $F_{\mu\nu}$, exactly the data required of the equations of motion.  Furthermore, we are to able to identify as equivalent any data that differ by a gauge transformation because the gauge invariance of the action ensures that it is well defined on the equivalence classes.  Thus we have a well-posed action principle. 

\subsection{Adding total derivatives}

Adding a boundary term such as the Gibbons-Hawking-York term amounts to adding a total derivative to the lagrangian.  We would like to understand how adding a total derivative can change the variational problem.   The bulk equations of motion are always unaffected by adding a total derivative, even one that contains higher derivatives of the fields.  However, the addition of a total derivative may render the variational problem inconsistent, or require different boundary data in order to remain well-posed.

Consider adding a total derivative to the lagrangian $L_1=\half \dot q^2$ of the free non-relativistic point particle, so that the action reads
\be S_2=\int _{t_1}^{t_2}dt\ L_2, \ \ L_2=-\half q\ddot q.\ee
The lagrangian contains higher derivatives, but they appear only as total derivatives, so the equations of motion are still second order.  This is analogous to the Einstein-Hilbert action without the GHY term. 

Varying the action produces the endpoint contribution
\be \half\left.\left(\dot q\delta q-q\delta \dot q\right)\right|_{t_1}^{t_2}.\ee
Setting $q(t_1)$ and $q(t_2)$, and the associated requirement $\delta q=0$ at the endpoints, is no longer sufficient to kill the surface contribution in the variation.   Fixing $\dot q(t_1)$ and $\dot q(t_2)$, as well as $q(t_1)$ and $q(t_2)$, and hence $\delta q=0$ and $\delta \dot q=0$, would be sufficient to kill the endpoint term.  But this is now setting 4 pieces of boundary data for a second order equation of motion, so for most choices of data the equations of motion will have no solution.  

One might try to say the following.  In the four dimensional space of variables that are fixed, parametrized by $q(t_1), q(t_2),\dot q(t_1),\dot q(t_2)$, there is some two dimensional subspace that is unconstrained.  The others are fixed by the equations of motion, so the parameters of this subspace are the true degrees of freedom of the theory.  In fact, there are many such subspaces, i.e. that parametrized by $q(t_1), q(t_2)$, or that parametrized by $q(t_1), \dot q(t_2)$, etc.  The essential point is that the action should be stationary when only variations of the quantities parametrizing the subspace are held fixed, because we must vary over all paths.  This is not true of the two subspaces just given.  If we could find a subspace for which this were true, we could salvage the action $S_2$, by saying that the degrees of freedom have been mixed up in the $q$'s and $\dot q$'s.  

For example, we might try writing the boundary contribution as 
\be -\half\left.q^2\delta\left(\dot q\over q\right)\right|_{t_1}^{t_2},\ee
and fixing the quantity ${\dot q\over q}$ on the boundary.  If fixing endpoint data for $\dot q/ q$ leads to a unique solution for most choices, then the endpoint variation vanishes upon fixing only this quantity, and we can say that the degree of freedom is encoded in the combination $\dot q/ q$. However, this does not work, since the quantities ${\dot q\over q}(t_1)$ and ${\dot q\over q}(t_2)$ do not parametrize a subspace over which the equation of motion is well-posed.  We can see this by noting that the general solution to the equation of motion is $q(t)=At+B$, so fixing ${\dot q\over q}(t_2)$ at the endpoints yields 
\be {A\over At_1+B}=C_1,\ \ \ {A\over At_2+B}=C_2,\ee
which cannot be solved for $A,B$ given generic values of $C_1,C_2,t_1,t_2$.  

In fact, there is no way to find such a subspace for the action based on $L_2$.\footnote{In general, to set up such a two dimensional subspace we must set functions of $q$ or $\dot{q}$ on the boundary separately, or a function $F(q,\dot{q})$ on both boundaries. It is clear that the first approach fails to set the above boundary variation to zero. We may imagine functions $F(q,\dot{q})$ and $f$ such that$f\delta F =\frac{1}{2}(\dot{q}\delta q-q\delta\dot{q})$. These would have the desired property that setting $F=C_{1}|_{t_{1}}$ and $F=C_{2}|_{t_{2}}$ parametrize a two dimensional subspace
and would imply the vanishing of the boundary variation. It is not difficult to show, however, that any such $F$ must be of the form $F(\frac{\dot{q}}{q})$ and thus is not a valid choice for the same reason that $\frac{\dot{q}}{q}$ is not.}   For this reason, if we are given the lagrangian $L_2$, we must add the GHY type term $\left.\left(\half q \dot q\right)\right|_{t_1}^{t_2}$, to bring the lagrangian back to $L_1$, which is well-posed.  As a general rule of thumb, if an action contains higher derivatives which appear only as total derivatives, a boundary term will need to be added.  

As a final example, which cleanly illustrates why the space of competing curves in the variational principle must be kept as large as possible, consider the lagrangian $L_3=\half \dot q^2+\dddot q.$ The boundary variation is then $\left.\dot q\delta q+\delta\ddot q\right|_{t_1}^{t_2}$.  We are fixing $\delta q$, and we might think we can simply fix $\ddot q=0$ as well.   As long as we can fix $\ddot q$ to be those values given by solving the equations of motion (i.e. $\ddot q=0$), it is still true that there is a two parameter ($q(t_1)$ and $q(t_2)$) family of endpoint data, all of which yield a unique solution to the equation of motion.  We are tempted to think of $\ddot q$ as a second-class constrained variable. However, fixing $\ddot q$ and hence $\delta\ddot q=0$ at the endpoints restricts the class of paths over which we are extremizing.  Taken to the extreme, we could also fix $\delta\dddot q=0$ at the endpoints, and so forth until the entire Taylor series is fixed and the only curve competing for the extremum is the solution itself.  To avoid arbitrariness, we need to keep the space of competing paths as large as possible by varying over all paths consistent with the specified boundary data.  The action based on $L_3$ is not well-posed.  

\subsection{\label{higherderivative} Higher derivative lagrangians}

We now turn to an example with higher derivatives that is analogous to the $F(R)$ gravity case.  Consider the action for a single variable $q(t)$ that involves an arbitrary function $F$ of the second derivative $\ddot q$, which satisfies $F''\not=0$, 
\be S_1=\int _{t_1}^{t_2}dt \ F(\ddot q),\ \ \ F''\not=0.\ee
Variation gives the following
\be \delta S_1=\int _{t_1}^{t_2}dt \ \left[F'''(\ddot q)\dddot q^2+F''(\ddot q)\ddddot q\right]\delta q+\left.\left[F'(\ddot q)\delta\dot q-F''(\ddot q)\dddot q\delta q\right]\right|_{t_1}^{t_2}.\ee
The bulk equation of motion can be solved for the highest derivative $\ddddot q$ in terms of the lower derivatives.  This is a fourth order equation in standard form, and requires four pieces of boundary data to be well-posed.  Fixing $q$ and $\dot q$ at both endpoints is a valid choice.  With this, the variation at the endpoints vanishes and we have a well defined action principle for two degrees of freedom,
\be S_1=S_1[q_1,\dot q_1,q_2,\dot q_2,t].\ee

Consider now introducing an auxiliary field, $\lambda(t)$, to try and get rid of the higher derivatives (this can always be done, a general method is the Ostrogradski method \cite{Ostrogradski,Nakamura:1995qz,Woodard:2006nt}),
\be S_2= \int _{t_1}^{t_2}dt \ F(\lambda)+F'(\lambda)(\ddot q-\lambda).\ee
Varying gives 
\be \delta S_2= \int _{t_1}^{t_2}dt \ \left[F'''(\lambda)\dot\lambda^2+F''(\lambda)\ddot\lambda\right]\delta q+F''(\lambda)\left[\ddot q-\lambda\right]\delta\lambda+\left.\left[F'(\lambda)\delta\dot q-F''(\lambda)\dot\lambda\delta q\right]\right|_{t_1}^{t_2}.\ee
This action is well-posed if $q$ and $\dot q$ are held fixed at the boundary, and $\lambda$ is kept arbitrary.  The equation of motion for $\lambda$ can be solved for $\lambda=\ddot q$, which when plugged back into the action yields $S_1$, and hence $S_1$ and $S_2$ are equivalent by the fundamental theorem of auxiliary variables (see Appendix \ref{fundamentaltheorem}).  

The action $S_2$ still involves the higher derivatives $\ddot q$, but now they appear only through a total derivative, so we integrate by parts,
\be S_2= \int _{t_1}^{t_2}dt \ F(\lambda)-\lambda F'(\lambda)-F''(\lambda)\dot q\dot\lambda+\left.F'(\lambda)\dot q\right|_{t_1}^{t_2}.\ee
The integration by parts has generated a boundary contribution.  We can render the action first order by subtracting this boundary term, that is, by adding to $S_2$ the GHY type term $\left.-F'(\lambda)\dot q\right|_{t_1}^{t_2}$.

The action and its variation now read
\be S_3= \int _{t_1}^{t_2}dt \ F(\lambda)-\lambda F'(\lambda)-F''(\lambda)\dot q\dot\lambda,\ee
\be \delta S_3= \int _{t_1}^{t_2}dt \ \left[F'''(\lambda)\dot\lambda^2+F''(\lambda)\ddot\lambda\right]\delta q+F''(\lambda)\left[\ddot q-\lambda\right]\delta\lambda-\left.\left[F''(\lambda)(\dot\lambda\delta q + \dot q\delta\lambda)\right]\right|_{t_1}^{t_2}.\ee
The bulk variation is unchanged because we have only added a total derivative, but the boundary variation has changed.  We must modify the variational principle to keep $\lambda$ and $q$ fixed on the boundary.  The action now becomes a functional $S_3=S_3[q_1,\lambda_1,q_2,\lambda_2,t]$.  The degree of freedom $\dot q$ has has been shifted into $\lambda$.  

Here adding the GHY type term was not strictly necessary to keep the variational principle well defined, as was the case in the previous section.  Here the effect is simply to shift the degrees of freedom into different variables.\footnote{This may be true of $F(R)$ theory as well, that is, the boundary term obtained from the correspondence to scalar-tensor theory is not necessarily the only one that renders the variational principle well-posed.  It is, however, the one that must be used if the correspondence to scalar-tensor theory is to be maintained at the boundary.  }

We are still free to eliminate the auxiliary variable $\lambda$ from the equations of motion.  Doing so yields the action
\be S_4=\int _{t_1}^{t_2}dt \ F(\ddot q)-{d\over dt}\left(F'(\ddot q)\dot q\right),\ee
with variation
\be \delta S_4=\int _{t_1}^{t_2}dt \ \left[F'''(\ddot q)\dddot q^2+F'(\ddot q)\ddddot q\right]\delta q-\left.\left[F''(\ddot q)\dddot q\delta q+F''(\ddot q)\dot q\delta \ddot q\right]\right|_{t_1}^{t_2}.\ee
The action is now a functional $S_4=S_4[q_1,\ddot q_1, q_2,\ddot q_2,t]$.  The degrees of freedom have been relabeled as $q$ and $\ddot q$.

The moral of all this is that while adding total derivatives (equivalently, boundary terms) to the action does not change the bulk equations of motion, it can change the variational principle, and the corresponding labeling of degrees of freedom, or it can make it impossible for the variational principle to be well-posed.  

\section{\label{gibbonshawking} Review of the Gibbons-Hawking-York term}

The GHY boundary term is a modification to the Einstein-Hilbert action which makes the action well-posed.  The modified action may be written as (conventions and definitions for boundary quantities laid out in appendices \ref{31appendix} and \ref{foliationappendix}),
\be
16\pi G\ S=16\pi G(S_{EH}+S_{GHY})=\int_{\mathcal{V}}d^{n}x\sqrt{-g}R + 2\oint_{\partial\mathcal{V}}d^{n-1}x\sqrt{|h|}K,
\ee
Where $G$ is Newton's constant.  Upon varying the action, we arrive at
\bea \nn
16\pi G\ \delta S=&&\int_{\mathcal{V}}d^{n}x\sqrt{-g}\left(R_{\mu\nu}-\frac{1}{2}Rg_{\mu\nu}\right)\delta g^{\mu\nu} -\oint_{\partial\mathcal{V}}d^{n-1}x\sqrt{|h|}h^{\alpha\beta}n^{\mu}\partial_{\mu}\delta g_{\alpha\beta}\\ &+& 2\oint_{\partial\mathcal{V}}d^{n-1}x\sqrt{|h|}\delta K.
\eea
Here we have used the assumption $\delta g^{\mu\nu} = 0$ on $\partial\mathcal{V}$, which also implies that the tangential derivative vanishes on $\partial\mathcal{V}$, $h^{\alpha\beta}\partial_\alpha \delta g_{\mu\beta}=0.$ Noting that
\begin{eqnarray} \nn
\delta K&=&\delta\left(h^{\alpha\beta}(\partial_{\alpha}n_{\beta}-\Gamma^{\mu}_{\alpha\beta}n_{\mu})\right)\\ \nn
&=&-h^{\alpha\beta}\delta\Gamma^{\mu}_{\alpha\beta}n_{\mu}\\
&=&\frac{1}{2}h^{\alpha\beta}n^{\mu}\partial_{\mu}\delta g_{\alpha\beta},
\end{eqnarray}
We see that
\be
16\pi G\ \delta S=\int_{\mathcal{V}}d^{n}x\sqrt{-g}\left(R_{\mu\nu}-\frac{1}{2}Rg_{\mu\nu}\right)\delta g^{\mu\nu}.
\ee
We now have the boundary variation vanishing without any restriction on the normal derivatives.  However, if this is the only property we desire, the choice of $2 K$ for the boundary term is not unique.  We are free to add an arbitrary function of the metric, normal vector, and tangential derivatives, $F(g_{\mu\nu},n_\mu,h^{\alpha\beta}\partial_{\beta})$, because the variation of such an addition vanishes with the assumption $\delta g_{\mu\nu} = 0$ on $\partial\mathcal{V}$.  

In fact, because of this freedom, Einstein unwittingly used the GHY boundary term well before either Gibbons, Hawking, or York proposed it \cite{Einstein:1916cd}.  He used an object $H$ for the lagrangian, instead of $R$, 
\be H=g^{\alpha\beta}\left(\Gamma^{\nu}_{\mu\alpha}\Gamma^{\mu}_{\nu\beta}-\Gamma^{\mu}_{\mu\nu}\Gamma^{\nu}_{\alpha\beta}\right).\ee
This is sometimes called the gamma-gamma lagrangian for GR, and has the advantage that it is first order in the metric, so there is no need to fix derivatives of the metric at the boundary.  

$H$ differs from $R$ by a total derivative,
\begin{eqnarray*}
H = R - \nabla_{\alpha}A^{\alpha},
\end{eqnarray*}
where $A^{\alpha} = g^{\mu\nu}\Gamma^{\alpha}_{\mu\nu}-g^{\alpha\mu}\Gamma^{\nu}_{\mu\nu}$. 
As such, it produces the same equations of motion, i.e. the Einstein equations, upon variation.  It also possesses all the same bulk symmetries as the Einstein-Hilbert action, namely diffeomorphism invariance.  Under a diffeomorphism, it does not change like a scalar, but it does change by a total derivative.

We can see that the gamma-gamma action differs from the Einstein-Hilbert plus GHY action by a boundary term of the form $F(g_{\mu\nu},n_\mu,h^{\alpha\beta}\partial_{\beta})$,
\begin{eqnarray} \nn
\int_{\mathcal{V}}d^{n}x\sqrt{-g}H&=&\int_{{\mathcal{V}}}d^{n}x\sqrt{-g}\left[R-\nabla_{\alpha}A^{\alpha}\right]\\
&=&\int_{\mathcal{V}}d^{n}x\sqrt{-g} R-\oint_{\partial\mathcal{V}}d^{n-1}x\ A^{\alpha}n_{\alpha}.
\end{eqnarray}
But\footnote{In detail, \beas
n_{\alpha}A^{\alpha}&=&n_{\alpha}(g^{\mu\nu}\Gamma^{\alpha}_{\mu\nu}-g^{\alpha\mu}\Gamma^{\nu}_{\mu\nu})\\
&=&\left(n^{\sigma}g^{\mu\nu}-n^{\mu}g^{\nu\sigma}\right)\frac{1}{2}\left[\partial_{\mu}g_{\nu\sigma}+\partial_{\nu}g_{\mu\sigma}-\partial_{\sigma}g_{\mu\nu}\right]\\
&=&\left(n^{\sigma}h^{\mu\nu}-n^{\mu}h^{\nu\sigma}\right)\frac{1}{2}\left[\partial_{\mu}g_{\nu\sigma}+\partial_{\nu}g_{\mu\sigma}-\partial_{\sigma}g_{\mu\nu}\right]\\
&=&n_{\alpha}h^{\mu\nu}\Gamma^{\alpha}_{\mu\nu}-\frac{1}{2}n_{\alpha}g^{\mu\alpha}h^{\nu\sigma}\partial_{\mu}g_{\nu\sigma}\\
&=&2n_{\alpha}h^{\mu\nu}\Gamma^{\alpha}_{\mu\nu}-n^{\mu}h^{\nu\sigma}\partial_{\nu}g_{\sigma\mu}\\
&=&-2K + 2h^{\alpha\beta}\partial_{\beta}n_{\alpha}-n^{\mu}h^{\nu\sigma}\partial_{\nu}g_{\sigma\mu}.
\eeas}
\be
A^{\alpha}n_{\alpha}= -2K + 2h^{\alpha\beta}\partial_{\beta}n_{\alpha}-n^{\mu}h^{\nu\sigma}\partial_{\nu}g_{\sigma\mu},
\ee
so this is an example of a choice of boundary term that differs from $K$ by a function $F$, namely $F=2h^{\alpha\beta}\partial_{\beta}n_{\alpha}-n^{\mu}h^{\nu\sigma}\partial_{\nu}g_{\sigma\mu}$.

The Einstein-Hilbert plus GHY action requires even fewer variables than those of the metric to be fixed on the boundary.  Only the induced metric $h_{ab}$ need be fixed.   To see this, we foliate in ADM variables with timelike hypersurfaces relative to ${\cal B}$, 
\bea \nn 16\pi G\ S_{EH}&=& \int_{\mathcal{V}}d^{n}x\sqrt{-g}R= \int_{\mathcal{V}}d^{n}x\ {\cal N}\sqrt{-\gamma}\left[^{(n-1)}R-{\cal K}_{ab}{\cal K}^{ab}+{\cal K}^2 \right. \\ &+&\left. 2\nabla_\alpha\left(r^\beta\nabla_\beta r^\alpha-r^\alpha\nabla_\beta r^\beta\right)\right].\eea
The total derivative term, when reduced to a surface term, cancels against the GHY term.  So the action is
\bea \nn 16\pi G(S_{EH}+S_{GHY})&=&\int_{\mathcal{V}}d^{n}x\sqrt{-g}R+2\oint_{\mathcal{B}}d^{n-1}z\sqrt{|h|}{\cal K} \\ &=& \int_{\mathcal{V}}d^{n}x\ {\cal N}\sqrt{-\gamma}\left[^{(n-1)}R-{\cal K}_{ab}{\cal K}^{ab}+{\cal K}^2\right].\eea
(Here we are ignoring total time derivatives and the GHY terms on the endpoints, but a similar cancellation will apply there.) There are no radial derivatives of the lapse or shift, so their variation need not be set to zero on the boundary.  Fixing the induced metric on the boundary is sufficient to render the action stationary.  

For most choices of an induced metric  on $\partial {\cal V}$, the Einstein equations should produce a unique solution in ${\cal V}$, up to diffeomorphisms that vanish at the boundary.  In this case, the Einstein-Hilbert plus GHY action is well-posed.  The counting in four dimensions goes as follows.  Of the ten pieces of boundary data for the ten components of the metric, there are 4 constraints.  The six components of the induced metric can be taken as the unconstrained components.  These are subject to equivalence under four gauge transformations, leaving two independent pieces of data, corresponding to the two degrees of freedom of the graviton. 

\section{\label{scalartensor}Higher derivative gravity and scalar-tensor theory}
Higher derivative theories of gravity have been studied extensively in many different contexts (for a review see \cite{Farhoudi:2005qd}). They are invoked to explain cosmic acceleration without the need for dark energy \cite{Boisseau:2000pr,Carroll:2003wy,Nojiri:2003ft,Woodard:2006nt,Carroll:2004de}, and as quantum corrections to Einstein gravity \cite{Asorey:1996hz,Bojowald:2006ww}.  

One can imagine trying to modify gravity by writing a quite general function of arbitrary curvature invariants, of any order in metric derivatives, 
\be  \int d^n x \ \sqrt{-g} F(R,R_{\mu\nu}R^{\mu\nu},R_{\mu\nu\lambda\sigma}R^{\mu\nu\lambda\sigma}, \ldots, \nabla_\mu R\nabla^\mu R ,R_{\mu\nu}R^{\mu}_{\ \lambda}R^{\lambda\nu},\ldots).\ee
Typically these are true higher derivative lagrangians, i.e. second and higher derivatives of the metric appear in a way that cannot be removed by adding total derivatives to the action, so the equations of motion are at least fourth order.  The amount of gauge symmetry is typically unchanged from general relativity, i.e.  diffeomorphism invariance remains the only gauge symmetry.  This means that the theories either have more degrees of freedom than general relativity, the presence of second class constraints/auxiliary fields, or both.  

Such models are not as diverse as it might seem, since they are essentially all equivalent to various multiple scalar-tensor theories \cite{Teyssandier:1983zz,Whitt:1984pd,Barrow:1988xh,Barrow:1988xi,Wands:1993uu,Chiba:2003ir,Chiba:2005nz}.  The transition to scalar-tensor theory amounts to an elimination of auxiliary fields represented by some of the higher derivatives of the metric.  They are  replaced by scalar fields that more efficiently encapsulate the physical degrees of freedom.  

We can get GHY terms for such theories by exploiting this equivalence.  The equivalent scalar-tensor theory is typically a Brans-Dicke like theory in Jordan frame with a potential for the scalar fields, minimally coupled to matter.  The theory can be brought to Einstein frame by a conformal transformation.  The boundary term in Einstein frame is just the GHY term, so we can find the boundary term for the original higher derivative theory by taking the Einstein frame GHY term backwards through the conformal transformation and through the scalar-tensor equivalence.  The term found in this way must be the correct one if the equivalence between the higher derivative theory and scalar-tensor theory is to hold for the boundary terms.  

The GHY terms obtained in this way are not generally sufficient to kill the boundary variation of the action when only $\delta g_{\mu\nu}=0$.  This is simply a reflection of the fact that we are dealing with a higher-order theory.  Some of the boundary values that the action depends upon involve derivatives of the metric, in a fashion exactly analogous to the fourth order toy example in section \ref{higherderivative}.  The only metric theories where $\delta g_{\mu\nu}=0$ should be sufficient are those with the special property that the equations of motion are still second order, despite the appearance of higher order terms in the action, namely the Lovelock lagrangians \cite{Lovelock:1971yv}.  Indeed, such GHY terms can be found for Lovelock theory \cite{Myers:1987yn,Charmousis:2005ey,Liu:2008zf}.  

In what follows, we analyze in detail the simplest case, namely the case where the lagrangian is allowed to be an arbitrary function of the Ricci scalar, $F(R)$, but does not contain any other curvature invariants.    The extension to more complicated cases follows easily when the scalar-tensor equivalence is known.  

\subsection{$F(R)$ gravity}

$F(R)$ theory is one of the most widely studied modifications of gravity \cite{Nojiri:2006ri,Sotiriou:2008rp}.  It has the ability to explain cosmic acceleration without dark energy \cite{Boisseau:2000pr,Carroll:2003wy,Nojiri:2003ft,Woodard:2006nt} and to evade local solar system constraints \cite{Nojiri:2003ft,Hu:2007nk}.    

The action for $F(R)$ gravity is 
\be S= \int d^n x \ \sqrt{-g}F(R).\ee
We would typically add matter which is minimally coupled to the metric, but it plays no essential role in the boundary terms, so we will omit the matter in what follows.  The Euler-Lagrange variation gives equations of motion which are fourth order in the metric.

The equivalence to scalar-tensor theory is seen by introducing a scalar field, $\phi$,  
\be S=\int d^n x \sqrt{-g}\left[F(\phi)+F'(\phi)(R-\phi)\right].\ee
The equation of motion for the scalar is 
\be F''(\phi)\left(R-\phi\right)=0,\ee
which, provided $F''\not= 0$, implies $R=\phi$.  Note that the scalar has mass dimension 2. Plugging this back into the action, using the fundamental theorem of auxiliary fields, recovers the original $F(R)$ action, so the two are classically equivalent.  This is $\omega=0$ Brans-Dicke theory with a scalar field $F'(\phi)$, and a potential.  

In the GR limit $F(R)\rightarrow R$, we have $F''\rightarrow 0$ so the transformation breaks down.  As the limit is taken, the scalar field decouples from the theory \cite{Olmo:2006eh}.

\subsection{Boundary terms for general scalar-tensor theory}

In this section we consider a general scalar-tensor action of the form 
\be
\label{generic}
S = \int_{\cal V} d^nx \, \sqrt{-g} \left(f(\phi) R
- {1 \over 2}\lambda(\phi) g^{\mu\nu}\partial_\mu \phi \partial_\nu \phi - U(\phi)\right).
\ee
We show that this should be supplemented by the boundary term 
\be \label{scalartensorGHY} S_{GHY}=2\oint_{\partial {\cal V}} d^{n-1}x\sqrt{|h|}f(\phi)K.\ee
We do this by showing that this reduces to the usual GHY term upon conformal transformation to the Einstein frame. 

Once this is done, the equations of motion for the metric are obtained by setting $\delta g_{\mu\nu}=0$ on the boundary, 
\bea
\label{gequation}
2f(\phi)\left(R_{\mu\nu}-\frac{1}{2}R g_{\mu\nu}\right)+2\Box f(\phi)g_{\mu\nu}-2\nabla_\mu\nabla_\nu f(\phi)=T_{\mu\nu}^\phi,
\eea
where
\be
\label{Tphi}
T_{\mu\nu}^\phi=\lambda(\phi)\left[\partial_\mu \phi \partial_\nu \phi-\frac{1}{2}g_{\mu\nu}(\partial \phi)^2 \right ]-g_{\mu\nu}U(\phi).
\ee
The equation of motion for the scalar field is obtained by setting $\delta \phi=0$ on the boundary,
\be
\label{phiequation}
\lambda(\phi)\Box\phi+{1\over 2}\lambda'(\phi)\left(\partial\phi\right)^2-U'(\phi)+f'(\phi)R=0.
\ee

We now proceed with the conformal transformation, keeping careful track of all boundary contributions \cite{Casadio:2001ff}.  Assuming
\[f(\phi)\not=0,\]
we can rewrite the action in terms of a conformaly re-scaled metric,
\be
\tilde{g}_{\mu\nu}=\left[16\pi Gf(\phi)\right]^{2\over n-2}g_{\mu\nu}, \ \ \ \begin{cases}G>0 \ \text{if}& f(\phi)>0,\\ G<0 \ \text{if} & f(\phi)<0.\end{cases}
\ee
$G$ can be chosen to be anything consistent with the sign of $f$, and will become the Einstein frame Newton's constant.  Re-writing the action is just a matter of using the conformal transformation formulae we have collected for convenience in appendix \ref{conformalformulae}.  In particular, we see from the second term of (\ref{rtildetransform}), used to rewrite $R$, that there is an integration by parts that will be necessary to bring the scalar kinetic term to its usual form.  This will generate a surface term which must be combined with the conformal transformation of the GHY term (\ref{scalartensorGHY}). 

The result is,  
 \be
\label{generic2}
S = \int_{\cal V} d^nx \, \sqrt{-\tilde{g}} \left({1\over 16\pi G} \tilde{R}
- {1 \over 2} A(\phi)\tilde{g}^{\mu\nu}\partial_\mu \phi \partial_\nu \phi - V(\phi)\right)+{1\over 8\pi G}\oint_{\partial {\cal V}} d^{n-1}x\sqrt{|\tilde h|}\tilde K,
\ee
where
\bea
&&A(\phi)={1\over 16\pi G}\left({1\over 2} {\lambda(\phi)\over f(\phi)}+{n-1\over n-2}{f'(\phi)^2\over f(\phi)^2}\right), \\
&& V(\phi)={U(\phi)\over \left[16\pi G f(\phi)\right]^{n\over n-2}}.
\eea
The GHY term has the usual form in Einstein frame, so working backwards, the Jordan frame expression must be correct as well.  Variation is done with $\delta g_{\mu\nu}=0$, $\delta\phi=0$ on the boundary.  

\subsection{Boundary term for $F(R)$ theory}

Adding the boundary term to the scalar-tensor form of the $F(R)$ action, we have 
\be S= \int_{\cal V} d^n x \sqrt{-g}\left[F(\phi)+F'(\phi)(R-\phi)\right]+2\oint _{\partial {\cal V}}d^{n-1}x\sqrt{|h|}F'(\phi)K.\ee
The variation of the action with respect to the scalar field now contains the boundary contribution
\be 2\oint_{\partial {\cal V}} d^{n-1}x\sqrt{|h|}F''(\phi)K\delta\phi,\ee
which vanishes since we require $\delta\phi=0$ on the boundary.  

Plugging back in we have 
\be S=\int _{\cal V} d^n x \ \sqrt{-g}F(R)+2\oint_{\partial {\cal V}}d^{n-1}x\sqrt{|h|}F'(R)K.\ee

This boundary term has been arrived at before in several different contexts, sometimes indirectly \cite{Balcerzak:2008bg,Barth:1984jb,Casadio:2001ff,Madsen:1989rz,Nojiri:1999nd}.  However, there has been some confusion, because this boundary term is not enough to make the action stationary given only $\delta g_{\mu\nu}=0$ on the boundary.  Indeed, there is in general no such boundary term with this property \cite{Madsen:1989rz}.  We see that this is because $R$ now carries the scalar degree of freedom, so we must set $\delta R=0$ on the boundary as well.  

We will now go on to accumulate some evidence that this is indeed the correct boundary term.  We first calculate the energy in the hamiltonian formalism, and show that to obtain the correct energy that reduces to the ADM energy when $F(R)\sim R$, this boundary term must be included.  We then calculate the entropy of Schwartzschild black holes, and show that in order to reproduce the results of the Wald entropy formula, the boundary term is necessary.  

\section{\label{hamiltonian} Hamiltonian formulation and ADM energy}
In this section, we will develop the hamiltonian formulation of a general scalar-tensor theory, which will encompass the $F(R)$ case.  We will stay in Jordan frame, and keep track of all boundary terms.  Just as in GR, the bulk hamiltonian vanishes on shell, and the only contribution comes from the boundary, so it will be essential to include the GHY terms found in the previous section.

We start with the action (\ref{generic}), which we write as $S=S_{G}+S_{\phi}+S_{B}$, where
\begin{eqnarray}
S_{G}&=&\int_{\mathcal{V}}d^{n}x \sqrt{-g} \ f(\phi)R,\\
S_{\phi}&=&\int_{\mathcal{V}}d^{n}x\sqrt{-g}\left[-\frac{1}{2}\lambda(\phi)g^{\alpha\beta}\partial_{\alpha}\phi\partial_{\beta}\phi - U(\phi)\right],\\
S_{GHY}&=&2\int_{\partial\mathcal{V}}d^{n-1}x\ \sqrt{ |h|} \  K f(\phi).
\end{eqnarray}

  We change to ADM variables \cite{Arnowitt:1962hi}, (see appendices \ref{31appendix} and \ref{foliationappendix} for conventions and definitions of the various quantities).  The boundary term splits into three integrals over the three boundaries.  The integral over $\Sigma_2$ gets an additional minus sign, since by convention the normal should be directed inward for spacelike surfaces, whereas for $\Sigma_2$ it is directed outward.  We will suppress the argument of $f$, $g$ and $\lambda$ over the course of the calculation.  
  
First look at the combination $S=S_{G}+S_{GHY}$,
\begin{eqnarray} \nn
S_{G}+S_{GHY}&=&\int_{\mathcal{V}}d^{n}x \sqrt{-g} \ f[^{(n-1)}R-K^{2}+K^{ab}K_{ab}-2\nabla_{\alpha}(n^{\beta}\nabla_{\beta}n^{\alpha}-n^{\alpha}K)]\\ \nn
&&+2\oint_{\mathcal{B}}d^{n-1}z\sqrt{-\gamma}\mathcal{K}f+2\oint_{\Sigma_{1}}d^{n-1}y\sqrt{h}Kf-2\oint_{\Sigma_{2}}d^{n-1}y\sqrt{h}Kf. \\
\end{eqnarray}
Integrating by parts the last term in the bulk integral, we find surface contributions that exactly cancel the boundary integrals over $\Sigma_1$ and $\Sigma_2$.  The surface contribution over ${\cal B}$ does not cancel its corresponding boundary integral, and we are left with,
\begin{eqnarray} \nn
S_{G}+S_{GHY}&=&\int_{\mathcal{V}}d^{n}x \sqrt{h} N \left[f\left(^{(n-1)}R-K^{2}+K^{ab}K_{ab}\right)+2f^{\prime}(n^{\beta}\nabla_{\beta}n^{\alpha}-n^{\alpha}K)\partial_{\alpha}\phi\right] \\
&&+2\oint_{\mathcal{B}}d^{n-1}z\sqrt{\gamma}f[\mathcal{K}-r_{\alpha}n^{\beta}\nabla_{\beta}n^{\alpha}].
\end{eqnarray}
We can simplify the integrand of the boundary piece,
\begin{eqnarray} \nn
\mathcal{K}-r_{\alpha}n^{\beta}\nabla_{\beta}n^{\alpha}&=&\mathcal{K}+n^{\alpha}n^{\beta}\nabla_{\beta}r_{\alpha}\\ \nn
&=&g^{\alpha\beta}\nabla_{\beta}r_{\alpha} +n^{\alpha}n^{\beta}\nabla_{\beta}r_{\alpha}\\ \nn
&=&(\sigma^{AB}e^{\alpha}_{A}e^{\beta}_{B}-n^{\alpha}n^{\beta})\nabla_{\beta}r_{\alpha}+n^{\alpha}n^{\beta}\nabla_{\beta}r_{\alpha}\\ \nn
&=&\sigma^{AB}e^{\alpha}_{A}e^{\beta}_{B}\nabla_{\beta}r_{\alpha}\\
&=&k.
\end{eqnarray}
The bulk terms multiplying $f^{\prime}$ can be further simplified\footnote{In detail, \begin{eqnarray*}
n^{\beta}\nabla_{\beta}n^{\alpha}&=&g^{\alpha\kappa}n^{\beta}\nabla_{\beta}n_{\kappa}\\
&=&(h^{\alpha\kappa}-n^{\alpha}n^{\kappa})n^{\beta}\nabla_{\beta}n_{\kappa}\\
&=&h^{\alpha\kappa}n^{\beta}\nabla_{\beta}n_{\kappa}\\
&=&h^{\alpha\kappa}n^{\beta}\partial_{\beta}n_{\kappa}-h^{\alpha\kappa}n^{\beta}n_{\lambda}\Gamma_{\beta\kappa}^{\lambda}\\
&=&h^{\alpha\kappa}n^{\beta}\partial_{\beta}n_{\kappa}-\frac{1}{2}h^{\alpha\kappa}n^{\beta}n^{\sigma}(\partial_{\beta}g_{\kappa\sigma} +\partial_{\kappa}g_{\beta\sigma}-\partial_{\sigma}g_{\beta\kappa})\\
&=&h^{\alpha\kappa}n^{\beta}\partial_{\beta}n_{\kappa}-\frac{1}{2}h^{\alpha\kappa}n^{\beta}n^{\sigma}\partial_{\kappa}g_{\beta\sigma}\\
&=&h^{\alpha\kappa}n^{\beta}\partial_{\beta}n_{\kappa}+\frac{1}{2}h^{\alpha\kappa}n^{\beta}n^{\sigma}\partial_{\kappa}(n_{\beta}n_{\sigma})\\
&=&h^{\alpha\kappa}n^{\beta}\partial_{\beta}n_{\kappa}-h^{\alpha\kappa}n^{\beta}\partial_{\kappa}n_{\beta}.\\
\end{eqnarray*}},
\be
n^{\beta}\nabla_{\beta}n^{\alpha}=h^{\alpha\kappa}n^{\beta}\left(\partial_{\beta}n_{\kappa}-\partial_{\kappa}n_{\beta}\right).
\ee

Putting all this together, we have
\begin{eqnarray} \nn
S_{G}+S_{GHY}&=&\int_{\mathcal{V}}d^{n}x \ \ \sqrt{h} N\left[ f\left(^{(n-1)}R-K^{2}+K^{ab}K_{ab}\right)\right. \\ \nn
&&+\left.2f^{\prime}\left(h^{\alpha\kappa}n^{\beta}(\partial_{\beta}n_{\kappa}-\partial_{\kappa}n_{\beta})-n^{\alpha}K\right)\partial_{\alpha}\phi\right] \\
&&+2\oint_{\mathcal{B}}d^{n-1}z\sqrt{\sigma}Nfk.
\end{eqnarray}
We now specialize to the $(t,y^{a})$ coordinate system, in which $n_{\alpha}=-N\delta^{0}_{\alpha}$, $e^{a}_{\alpha}=\delta^{a}_{\alpha}$.  The term $h^{\alpha\kappa}n^{\beta}\partial_{\beta}n_{\kappa}$ vanishes.  We are left with
\begin{eqnarray} \nn
S_{G}+S_{GHY}&=&\int_{\mathcal{V}}d^{n}x \sqrt{h}\left[N f\left(^{(n-1)}R-K^{2}+K^{ab}K_{ab}\right)\right. \\ \nn &&+\left.2f^{\prime}(h^{ab}\partial_{a}N\partial_{b}\phi-K\dot{\phi} + KN^{a}\partial_{a}\phi)\right]\\
&&+2\oint_{\mathcal{B}}d^{n-1}z\sqrt{\sigma}Nfk.
\end{eqnarray}

The scalar action, in ADM variables, is
\be S_\phi=\int_{\mathcal{V}}d^{n}x\ {\sqrt{h}\lambda\over 2N}\left[\dot\phi\left(\dot\phi-2N^a\partial_a\phi\right)-N^2h^{ab}\partial_a\phi\partial_b\phi+(N^a\partial_a\phi)^2\right]-N\sqrt{h}U.
\ee

Define now
\begin{eqnarray}
S^{\prime}_{G}&\equiv&\int_{\mathcal{V}}d^{n}x \sqrt{h}\left(N f[^{(n-1)}R-K^{2}+K^{ab}K_{ab}]+2f^{\prime}h^{ab}\partial_{a}N\partial_{b}\phi\right),\\
S^{\prime}_{\phi}&\equiv&S_{\phi}-2\int_{\mathcal{V}}d^{n}x \sqrt{h}f^{\prime}K(\dot{\phi}-N^{a}\partial_{a}\phi),\\
S^{\prime}_{B}&\equiv&2\oint_{\mathcal{B}}d^{n-1}z\sqrt{\sigma}Nfk,
\end{eqnarray}
\be S=S^{\prime}_{G}+S^{\prime}_{\phi}+S^{\prime}_{B}. \ee

The action is now in a form amenable to transition to the hamiltonian, namely it is in the form of a time integral over a lagrangian, $L$, (which is itself a space integral plus boundary parts) containing no time derivatives higher than first, and no boundary contributions at the time endpoints,
\be S=\int_{t_1}^{t_2}dt\ L\left[h_{ab},\dot h_{ab},N,N^a,\phi,\dot\phi\right].\ee
Note that this would not be the case were it not for the GHY term on the surfaces $\Sigma_1$ and $\Sigma_2$.  It now remains to transition to the hamiltonian formulation.  This has been done without keeping surface terms by \cite{Garay:1992ej,Capozziello:2007gm}, and at the level of the equations of motion by \cite{Salgado:2005hx}.  We start by finding the canonical momentum conjugate to $\phi$,
\be \label{momentaphi}
p_{\phi}=\frac{\delta{L}}{\delta\dot{\phi}}={\sqrt{h}\lambda\over N}\left(\dot\phi-N^a\partial_a\phi\right)-2\sqrt{h}f'K.
\ee

To find the canonical momenta conjugate to $h_{ab}$, we vary with respect to $K_{ab}$, then use the relation $K_{ab}=\frac{1}{2N}\left(\dot{h}_{ab}-\nabla_{a}N_{b}-\nabla_{b}N_{a}\right)$ to replace $\delta K_{ab}=\frac{1}{2N}\delta\dot{h}_{ab}$,
\bea \nn \label{momentak} p^{ab}&=&\frac{\delta{L}}{\delta \dot{h}_{ab}} \\ \nn &=&\sqrt{h}\left[f\left(K^{ab}-Kh^{ab}\right) -\frac{f^{\prime}}{N}h^{ab}(\dot{\phi}-N^{c}\partial_{c}\phi)\right].\\
\eea
Equations (\ref{momentaphi}) and (\ref{momentak}) are invertible for $\dot \phi$ and $K^{ab}$ (which is essentially $\dot h_{ab}$) in terms of $p_\phi$, $p^{ab}$ (and $\phi$, $h_{ab}$, $N$, $N^a$), 
\bea \label{momentummap} \dot\phi&=&{N\over \sqrt{h}}\left[(n-2)fp_\phi-2f'p\over 2(n-1)f'^2+(n-2)f\lambda\right]+N^a\partial_a\phi,\\ \nn K_{ab}&=&{1\over f}{p_{ab}\over \sqrt{h}}-h_{ab}\left[p_\phi f'+2p{f'^2\over f}+p\lambda\over 2(n-1)f'^2+(n-2)f\lambda\right],
\eea
where $p=h^{ab}p_{ab}$.  The canonical momenta conjugate to $N$ and $N^a$ both vanish, just as in GR. 

Notice that the map from velocities ($\dot\phi$, $\dot h_{ab}$) to momenta ($p_\phi$, $p_{ab}$) is non-singular even when the scalar kinetic term vanishes, $\lambda\rightarrow 0$, as in the case of theories equivalent to $F(R)$.  This corresponds to the fact that the scalar is still dynamical in this limit, by virtue of its non-minimal coupling.  The case $f\rightarrow const\not=0$ is also well behaved, provided $\lambda\not=0$, as the scalar has dynamics stemming from the kinetic term.  The case $f\rightarrow const$ and $\lambda\rightarrow 0$ is indeed singular, because the scalar field then loses its dynamics.  

We now start the calculation of the hamiltonian.  Starting with the scalar field part, we must express everything in terms of the fields and momenta, by eliminating all time derivatives, 
\begin{eqnarray} \nn
{H}^{\prime}_{\phi}&=&\left(\int_{\Sigma_{t}}d^{n-1}y \ p_{\phi}\dot{\phi}\right)-{L}^{\prime}_{\phi}\\ \nn
&=&\int_{\Sigma_{t}}d^{n-1}y \ \sqrt{h}\left[N\left(\frac{1}{2\lambda}\frac{p_{\phi}^2}{h} +2f^{\prime}\frac{K}{\lambda}\frac{p_{\phi}}{\sqrt{h}}+2{f^{\prime}}^{2}\frac{K^{2}}{\lambda}+\frac{1}{2}\lambda h^{ab}\partial_{a}\phi\partial_{b}\phi+U\right)\right. \\ &&+\left.N^{a}\left(\frac{p_{\phi}}{\sqrt{h}}\partial_{a}\phi\right)\right].
\end{eqnarray}
Here we must treat $K_{ab}$ as the function of $p^{ab}$, $h_{ab}$, $\phi$ and $p_{\phi}$ given by (\ref{momentummap}).  Naively, the above appears singular in the limit $\lambda\rightarrow 0$ (the $F(R)$ case), however once we re-express $K_{ab}$ in terms of the momenta, the $1/\lambda$ terms cancel and the limit is smooth.   

Now the metric part, 
\bea \nn
{H}^{\prime}_{G}&=&\left(\int_{\Sigma_{t}}d^{n-1}y \ p^{ab}\dot{h}_{ab}\right) -{L}^{\prime}_{G}\\ \nn
&=&\int_{\Sigma_{t}}d^{n-1}y \ p^{ab}(2NK_{ab}+2\nabla_{a}N_{b})-\sqrt{h}\left(N f[^{(n-1)}R-K^{2}+K^{ab}K_{ab}] \right. \\ &&\left.+2f^{\prime}h^{ab}\partial_{a}N\partial_{b}\phi\right).
\eea
Integrate by parts to pull all the derivatives off of $N$ and $N^a$, being sure to keep the boundary contributions,
\bea \nn
{H}^{\prime}_{G}&=& \int_{\Sigma_{t}}d^{n-1}y\ \sqrt{h}\left[2N{p^{ab}\over \sqrt{h}}K_{ab}-Nf\left(^{(n-1)}R-K^{2}+K^{ab}K_{ab}\right)\right. \\ \nn &&+\left.2N\nabla^a(f'\nabla_a\phi)-2N_a\nabla_b\left(p^{ab}\over \sqrt{h}\right)\right] \\
&&+2\oint_{S_{t}}d^{n-2}\theta\sqrt{\sigma}\ r_a\left(N_b{p^{ab}\over \sqrt{h}}-Nf'\partial^a\phi\right).
\eea

The boundary term in the lagrangian contributes 
\begin{eqnarray*}
{H}^{\prime}_{B}=-{L}^{\prime}_{B}=-2\oint_{S_t}d^{n-2}\theta\sqrt{\sigma}Nfk.
\end{eqnarray*}
Combining these terms yields the full hamiltonian
\begin{eqnarray} \nn
H&=&H'_G+H'_\phi+H'_B\\ \nn
&=&\int_{\Sigma_{t}}d^{n-1}y\sqrt{h}\left[Nf\left(-^{(n-1)}R-K^{2}+K^{ab}K_{ab}\right)\right.\\ \nn
&&+N\left(\frac{p^{ab}}{\sqrt{h}}K_{ab}+2\nabla^{a}(f^{\prime}\nabla_{a}\phi) +\frac{1}{2\lambda}\frac{p_{\phi}^2}{h}+{f^{\prime}}\frac{K}{\lambda}\frac{p_{\phi}}{\sqrt{h}}+\frac{1}{2}\lambda h^{ab}\partial_{a}\phi\partial_{b}\phi +U\right)\\ \nn
&&+\left.N_{a}\left(\frac{p_{\phi}}{\sqrt{h}}\partial^{a}\phi-2\nabla_{b}\frac{p^{ab}}{\sqrt{h}}\right)\right]\\
&&+2\oint_{S_{t}}d^{n-2}\theta\sqrt{\sigma}\left[r_{a}N_{b}\frac{p^{ab}}{\sqrt{h}}-N(fk+r^{a}f^{\prime}\partial_{a}\phi)\right].
\end{eqnarray}

This hamiltonian is like that for GR in the sense that the equations of motion for $N$ and $N^{a}$ are Lagrange multipliers which cause the bulk terms to vanish.  We have not bothered to write out $K_{ab}$ using (\ref{momentummap}) because it because it does not contain $N$ or $N^{a}$.  

The hamiltonian evaluated on solutions reduces to the boundary part
\be
H_{solution}=2\oint_{S_{t}}d^{n-2}\theta\sqrt{\sigma} r_aN_b{p^{ab}\over \sqrt{h}}-N(fk+r^af'\partial_a\phi).
\ee
The ADM energy, $E$, is given by choosing the lapse to vanish and the shift to be unity,
\be
E=-2\oint_{S_{t}}d^{n-2}\theta\ \sqrt{\sigma}\left[fk+f^{\prime}r^{a}\partial_{a}\phi\right].
\ee
Alternatively, we could also have obtained this expression by finding the hamiltonian in Einstein frame and then performing a conformal transformation.  

As an example, consider the Schwartzschild solution in four dimensions,
\be ds^2=-f(r)dt^2+{1\over f(r)}dr^2+r^2d\omega^2,\ \ \ \phi=\phi_0,\ee
where $f(r)=1-{2GM\over r}$, $GM$ is a constant, and $\phi_0$ is a constant.  This will be a solution to the scalar-tensor theory equations of motion (\ref{gequation}), (\ref{Tphi}), and (\ref{phiequation}), provided that Minkowski space is a solution, and $\phi_0$ is set to the vacuum value of $\phi$ in the Minkowski solution.  The ADM energy for this solution is\footnote{Recall that we must measure the energy relative to the energy of Minkowski space, which is the vacuum solution.  The energy of each is individually divergent as $r\rightarrow\infty$, but the difference is finite and yields the above expression.}
\be E=16\pi GM f(\phi_0),\ee
which is what one expects, given that $ f(\phi_0)$ is playing the role of the effective gravitational constant, $ f(\phi_0)={1\over 16\pi G_{\rm eff}}$.

$F(R)$ theory is the special case where $f(\phi)=F'(\phi)$, $U(\phi)=F'(\phi)\phi-F(\phi)$, $\lambda(\phi)=0$.  In order to hamiltonize $F(R)$ theory, it must first be brought to first order form (this is essentially the content of the Ostrogradski method for hamiltonizing higher order systems  \cite{Ostrogradski,Nakamura:1995qz,Woodard:2006nt}).  Passing to the scalar-tensor description is the simplest way to do this, so we have also found the ADM energy for $F(R)$ theory.  The boundary term must be passed in the same way, and so plays the same essential role.  

\section{\label{blackholes} Black hole entropy}

When calculating the entropy of a black hole in the euclidean semi-classical approximation, it is essential to have the correct GHY term \cite{Brown:1992bq}.  In fact, this term is responsible for the entire contribution to the euclidean action.  In this section we will calculate the entropy for a Schwartzschild black hole in $F(R)$ theory using our boundary term, and compare this to the entropy given by the Wald entropy formula.

The Wald entropy formula allows one to calculate the entropy of a black hole in any diffeomorphism invariant metric theory of gravity.  It involves an integral over the bifurcation two sphere of the horizon of the black hole \cite{Wald:1993nt,Iyer:1994ys}.  The formula does not rely on having a well-posed action principle, i.e. it depends only on the lagrangian density, and hence the GHY terms are not needed in order to apply it.  For a Schwartzschild black hole in $F(R)$ theory in four dimensions, the Wald formula gives the result \cite{Briscese:2007cd}
\be S_{BH}={A\over 4}16\pi F'(R_0),\ee
where $A$ is the area of the horizon and $R_0$ is the (constant) background curvature of the spacetime the black hole sits in.  Wald has shown that his entropy formula gives the same value as the euclidean semi-classical approach for theories which satisfy several conditions in addition to those needed by the formula itself \cite{Iyer:1995kg}.  One of these additional conditions is that there be a variational principle for the action where only the metric is held fixed on the boundary, so this does not cover the general $F(R)$ case.  Nevertheless, we will see that our surface term still gives the entropy in agreement with Wald's formula.  

The partition function in the semiclassical limit is
\be Z[\beta]=e^{-S_E},\ee
where $\beta={1\over T}$ is the inverse temperature and $S_E$ is the euclidean action of the dominant classical field configuration where the time variable is identified with period $\beta$.  

In Einstein gravity in four dimensions, we have the Schwartzschild solution 
\be ds^2=-f(r)dt^2+{1\over f(r)}dr^2+r^2d\omega^2,\ee
where $f(r)=1-{2GM\over r}$, and $M$ is the ADM mass.  

The temperature can be determined by finding the corresponding solution to the euclidean action, which amounts to taking $t\rightarrow i\tau$, and then finding the period of $\tau$ required to eliminate the conical singularity at the horizon.  The resulting metric is then
\be  ds_E^2=f(r)d\tau^2+{1\over f(r)}dr^2+r^2d\omega^2,\ \ \ \tau=\tau+\beta,\ \ \ \beta=8\pi GM. \ee
The temperature of a black hole is model-independent, since no use of the Einstein equations was made other than the fact that Schwartzschild is a solution.  Thus in any theory of modified gravity where Schwartzschild is a solution, the black hole will have the same temperature.  

We first review the euclidean action calculation in GR, since the extension to $F(R)$ is then trivial.  The euclidean action for GR is 
\be 16\pi GS_E=-\int_{\cal V} d^4x\ \sqrt{|g|}R-2\oint_{\partial{\cal V}} d^3x\ \sqrt{|h|}(K-K_0).\ee
The action for a general theory must be zeroed on the action for the background one is expanding about, i.e. the vacuum state.  The true, finite action is thus $S-S_0$, where $S_0$ is the action of the background.  In our case, the background is flat space.  The bulk contribution to $S_0$ vanishes, and all that remains is $K_0$, the extrinsic curvature of the boundary as measured in flat space.  The term $K_0$ is often called the boundary counterterm (misappropriated from the quantum field theory jargon) and may in general have to take a more complicated form \cite{Mann:2005yr}).

Since the Ricci curvature of Schwartzschild vanishes, the bulk term does not contribute to the classical action.  The entire contribution comes from the boundary term.  Taking the boundary to be a sphere of radius $r$ about the origin, we have 
\be \oint_{\partial{\cal V}}d^3x\ \sqrt{|h|}K=4\pi \beta(2r-3GM).\ee
For the background, we periodically identify the time to the period $\beta$, redshifted by the factor $\left(1-{2GM\over r}\right)^{1/2}$,
\be \oint_{\partial{\cal V}}d^3x\ \sqrt{|h|}K_0=8\pi \beta r\left(1-{2GM\over r}\right)^{1/2}.\ee
Taking the difference and the limit $r\rightarrow \infty$ yields
\be S_E={\beta^2\over 16\pi G}.\ee
The free energy, entropy, and energy are then
\bea F&=&-{1\over \beta}\ln Z={1\over \beta}S_E={\beta\over 16\pi G},\\
E&=&F+\beta{\partial F\over \partial \beta}={\beta\over 8\pi G}=M,\\
S&=& \beta^2{\partial F\over \partial \beta}={\beta^2\over 16\pi G}={A\over 4G},
\eea
where $A=4\pi(2GM)^2$ is the area of the horizon.   

We now extend the calculation to $F(R)$ theory.  The bulk vacuum equations of motion are 
\be F'(R)R_{\mu\nu}-\half F(R)g_{\mu\nu}+g_{\mu\nu}\nabla^2 F'(R)-\nabla_\mu\nabla_\nu F'(R)=0,\ee
which are fourth order in the metric, as expected.  AdS/dS Schwartzschild has the property $R_{\mu\nu}={R\over 4}g_{\mu\nu}$, with $R=R_0$ a constant.  Using this ansatz the equation of motion reduces to 
\be \label{frvacuum} F'(R_0)R_0-2 F(R_0)=0,\ee
which is an algebraic equation for $R_0$.  This is the same equation one would obtain seeking constant curvature solutions, so for every constant curvature background, we also have a Schwartzschild black hole that asymptotically approaches this background.  

The euclidean action, including the GHY term, is
\be S_E+S_0=-\int_{\cal V} d^4 x \ \sqrt{|g|}F(R)-2\oint_{\partial{\cal V}} d^{3}x\sqrt{|h|}F'(R)K.\ee
Here $S_0$ is the action for the background.  Since the Schwartzschild solution has the same constant curvature as the background, the bulk contribution vanishes, and the entire contribution again comes from the boundary term,
\be S_E=-2F'(R_0)\oint_{\partial{\cal V}} d^{3}x\sqrt{|h|}\left(K-K_0\right).\ee
where $R_0$ is the scalar curvature of the background/black hole.  

If we assume for simplicity that $F(R)$ is such that there is a flat space solution, then the calculation proceeds just as in the GR case, with the result
\be S_E={\beta^2}F'(R_0).\ee
The thermodynamic quantities are then
\be F={\beta}F'(R_0),\ \ \ E=16\pi G F'(R_0)M,\ \ \ S={A\over 4}16\pi F'(R_0).\ee

These formulae make good sense from the point of view of the scalar-tensor theory.  The scalar field is just $R$, the Ricci curvature, so the Schwartzschild solutions we are considering have constant scalar field everywhere.  This is in accord with no hair theorems, which forbid non-trivial scalar profiles around black holes in space with zero or positive curvature \cite{Bekenstein:1971hc,Heusler:1994wa,Winstanley:2005fu}.  The value of the effective Newton's constant is set by the constant value of the scalar field
\be {1\over 16\pi G_{eff}}=F'(R_0),\ee
so the entropy is simply one quarter the area of the horizon in units of the effective Planck length.  The energy is the ADM energy we found in the previous section.

An interesting consequence of this formula is that higher curvature terms make no correction to the entropy.  From (\ref{frvacuum}) we see that if $R=0$ is to be a solution, then $F(0)=0$, so $F$ can be taylor expanded around the origin, starting with the Einstein-Hilbert term: $F(R)=F'(0)R+\half F''(0)R^2+\cdots$.  All the higher power corrections to the action do not affect the entropy.  In particular, the entropy of a black hole in pure $R^n$ gravity, for $n\geq 2$, vanishes.  

\section{\label{conclusion} Conclusions}

Having correct boundary terms is essential to the consistency of any theory.  One can generally get away without them, but there are instances where they are vitally important.  We have argued that consistent GHY terms for higher derivative modified gravity theories can be obtained by using any scalar-tensor equivalence the theory may posses.  The boundary terms obtained in this way, while not necessarily unique, do give a well-posed variational problem and give the expected ADM energy and black hole entropy, even though derivatives of the metric may have to be fixed on the boundary.  

What we have given is by no means a complete analysis of boundary terms, however.  For a general lagrangian of any kind of field, not necessarily a modified gravity theory, it is far from clear whether there always exists a boundary term that renders the variational principle well-posed.  Furthermore, even if such a term can be found, it is not evident what freedom is allowed in choosing the boundary term, or what physical significance this freedom entails. 

\bigskip
\goodbreak
\centerline{\bf Acknowledgements}
\noindent
The authors are grateful to Allan Blaer, Adam Brown, Norman Christ, Solomon Endlich, Brian Greene, Dan Kabat, Alberto Nicolis, and Erick Weinberg for discussions, and to their anonymous referee for helpful comments.  KH is supported by DOE grant DE-FG02-92ER40699 and by a Columbia University Initiatives in Science and Engineering grant.  

\appendix

\section{\label{31appendix} The ADM decomposition}

Here we review the ADM hypersurface decomposition of spacetime, and lay out our conventions in the process.  The conventions and notation are those of \cite{Poisson}.  In this appendix we will describe a generic foliation of a volume ${\cal V}$, by spacelike or timelike hypersurfaces, and in the next appendix we will describe the specific foliations we use throughout the paper.  

Put coordinates $x^\mu$ on ${\cal V}$.  We foliate the volume ${\cal V}$ with hypersurfaces $\Sigma_t$ by giving a global time function $t(x^\mu)$ and declaring the hypersurfaces to be its level sets.  We then set up functions $y^a(x^\mu)$, $a=1\ldots n-1$, independent of $t(x^\mu)$ and each other, to serve as coordinates on the submanifolds.  Taken together, $t$ and $y^a$ form a new coordinate system on ${\cal V}$, and we have an invertible transformation from the these coordinates to the old ones 
\be x^\mu(y^a,t),\ \ \ \ \ t(x^\mu),\ y^a(x^\mu).\ee

The coordinate basis vectors of this new coordinate system are
\be t^\mu={\partial x^\mu\over \partial t},\ \ \ e^\mu_{\ a}={\partial x^\mu\over \partial y^a}.\ee
The dual one forms are 
\be \tilde t_\mu={\partial t\over \partial x^\mu},\ \ \ \tilde{e}^{\ a}_\mu={\partial y^a \over \partial x^\mu}.\ee
They satisfy the duality and completeness relations
\be t^\mu \tilde t_\mu=1,\ \ e^\mu_{\ b}\tilde{e}^{\ a}_\mu=\delta^a_b,\ \ t^\mu \tilde{e}^{\ a}_\mu=e^\mu_{\ a}\tilde t_\mu=0.\ee
\be t^\mu \tilde t_\nu+e^\mu_{\ a}\tilde{e}^{\ a}_\nu=\delta^\mu_{\ \nu}.\ee

Introduce a bulk metric $g_{\mu\nu}$.  There is now a well defined one-dimensional normal subspace at each point of $\Sigma_t$, which may be different from the subspace spanned by $t^\mu$.  We set up a basis consisting of the forward pointing unit normal vector $n^\mu$ along with the $e^\mu_{\ a}$. The $e^\mu_{\ a}$ are not required to be orthonormal among themselves, but are orthogonal to $n^\mu$.  We have
\be g_{\mu\nu}n^\mu n^\nu=\epsilon,\ \ g_{\mu\nu}e^\mu_{\ a}n^\nu=0. \ee
Here $\epsilon$ is defined by 
\be \epsilon=\begin{cases} 1& \Sigma_t \text{ timelike}\\ -1& \Sigma_t \text{ spacelike}\end{cases}.\ee

 We define the associated dual forms ${e}^{\ a}_\mu$, $\tilde n_\mu$, at each point, 
\be n^\mu \tilde n_\mu=1,\ \ e^\mu_{\ b}{e}^{\ a}_\mu=\delta^a_b,\ \ n^\mu {e}^{\ a}_\mu=e^\mu_{\ a}\tilde n_\mu=0.\ee
\be n^\mu \tilde n_\nu+e^\mu_{\ a}{e}^{\ a}_\nu=\delta^\mu_{\ \nu}.\ee
(We use $\tilde n_\mu$ for the dual one form, and reserve $n_\mu$ for the form $g_{\mu\nu}n^\nu$.  They differ by a sign for spacelike hypersurfaces, i.e. $\tilde n_\mu=\epsilon n_\mu$.)

A vector field $A^\mu$ is called parallel if it admits the decomposition $A^\mu=A^a e^\mu_{\ a}$.  A form $A_\mu$ is parallel if it admits the decomposition $A_\mu=A_a {e}_\mu^{\ a}$.   Similarly, a general tensor is parallel if it admits a similar decomposition, for example, a $(2,1)$ tensor is parallel if
\be A^{\mu\nu}_{\ \ \gamma}=A^{ab}_{\ \ c}e^\mu_{\ a}e^\nu_{\ b} {e}_\gamma^{\ c}.\ee
There is a bijective relation between tensors on the submanifold $\Sigma_t$ (really a one parameter family of tensors, one on each surface, parametrized by $t$) and parallel tensors in the bulk.  Given a parallel bulk tensor $A^{\mu\nu}_{\ \ \gamma}$, it corresponds to the submanifold tensor $A^{ab}_{\ \ c}$, and vice versa.  

Define the projection tensor
\be {P}^\mu_{\ \nu}=\delta^\mu_{\nu}-n^\mu\tilde n_{\nu}.\ee
It projects the tangent space of ${\cal V}$ onto the tangent space of $\Sigma_t$, along the subspace spanned by $n^\mu$.  It satisfies 
\bea {P}^\mu_{\ \lambda} {P}^\lambda_{\ \nu}={P}^\mu_{\ \nu},\\ {P}^\mu_{\ \nu}e^\nu_{\ a}=e^\mu_{\ a},\ \ {P}^\mu_{\ \nu}n^\nu=0 \\ {P}^\mu_{\ \nu}{e}_\mu^{\ a}={e}_\nu^{\ a}, \ \ P^\mu_{\ \nu}\tilde n_\mu=0.\eea

Given any bulk tensor, e.g. $A^{\mu\nu}_{\ \ \gamma}$, we can make a parallel tensor by projecting it,
\be A^{\parallel \mu\nu}_{\ \ \ \gamma}={P}^\mu_{\ \rho}{P}^\nu_{\ \sigma}{P}^\lambda_{\ \gamma} A^{\rho\sigma}_{\ \ \lambda}.\ee
A tensor is parallel if and only if it is equal to its projection.  

We have the relation
\be e^\mu_{\ a}{e}^{\ a}_{\nu}={P}^\mu_{\ \nu}.\ee  

Projecting the metric gives the induced metric $h$ on the hypersurfaces,
\be h_{\mu\nu}=P^\rho_{\ \mu}P^\sigma_{\ \nu}g_{\rho\sigma}=h_{ab}e_\mu^{\ a}e_\nu^{\ b},\ \ \ h_{ab}=e^\mu_{\ a}e^\nu_{\ b}h_{\mu\nu}=e^\mu_{\ a}e^\nu_{\ b}g_{\mu\nu}.\ee
We raise and lower bulk indices $\mu,\nu\ldots$ with $g_{\mu\nu}$ and its inverse $g^{\mu\nu}$, and we raise and lower submanifold indices $a,b,\ldots$ with $h_{ab}$ and its inverse $h^{ab}$.  In particular, we have, 
\bea h^{ab}g_{\mu\nu}e^\nu_{\ b}&=&e_\mu^{\ a},\ \ \ g_{\mu\nu}n^\nu\equiv n_\mu=\epsilon \tilde n_\mu,\\
h_{ab}g^{\mu\nu}e_\nu^{\ b}&=&e^\mu_{\ a},\ \ \ g^{\mu\nu}\tilde n_\nu\equiv \tilde n^\mu=\epsilon n^\mu ,\eea
as well as
\be g_{\mu\alpha}P^\alpha_{\ \nu}=h_{\mu\nu},\ \ \  g^{\mu\alpha}P^\nu_{\ \alpha}=h^{\mu\nu}.\ee

To perform the ADM decomposition, we want to write the bulk metric in the $(y^a,t)$ coordinate system.  We start by expanding the coordinate vector $t^\mu$ over the new basis,
\be t^\mu=Nn^\mu+N^{a}e^\mu_{\ a}.\ee
The coefficients $N$ and $N^a$ are called the lapse and the shift, respectively.  We have
\be \tilde t_\mu=\partial_\mu t= \epsilon{1\over N}n_\mu,\ \ \ n_\mu=\epsilon N \tilde t_\mu=\epsilon N\partial_\mu t.\ee

The coordinate one-forms are 
\be dx^\mu=t^\mu dt+e^\mu_{\ a}dy^a=Nn^\mu dt+\left( N^a dt+dy^a\right)e^\mu_{\ a}.\ee
The metric is
\be g_{\mu\nu}dx^\mu dx^\nu= \epsilon N^2 dt+h_{ab}\left(N^a dt +dy^a\right)\left(N^b dt +dy^b\right).\ee
In matrix form this is,
\be g= \left(\begin{array}{c|c}\epsilon N^2+N^a N_{a}  & N_{a} \\ \hline N_{a} & h_{ab}  \end{array}\right).\ee
The inverse metric is
\be g^{-1}= \left(\begin{array}{c|c} \epsilon{1\over N^2} & -\epsilon{N^a\over N^2} \\ \hline -\epsilon{N^a\over N^2} & h^{ab}+ \epsilon{N^a N^b\over N^2}\end{array}\right).\ee
The square root of the norm of the determinant is
\be \sqrt{|g|}=N\sqrt{|h|}.\ee
The normal vector is
\be n^0={1\over N},\ \ \ n^a=-{N^a\over N},\ \ \ n_0= \epsilon N,\ \ \ n_a=0.\ee

The extrinsic curvature is a parallel tensor defined by
\be K_{ab}\equiv  e^\mu_{\ a} e^\nu_{\ b}\nabla_\mu n_\nu.\ee
It is symmetric,
\be K_{ab}=K_{ba},\ee
as can be easily shown by noting that the basis vectors have zero Lie bracket, $e^\nu_{\ b}\nabla_\nu e^\mu_{\ a}=e^\nu_{\ a}\nabla_\nu e^\mu_{\ b}.$  We also have
\be K_{ab}=\nabla_{(\mu}n_{\nu)}e^\mu_{\ a}e^\nu_{\ b}=\half e^\mu_{\ a}e^\nu_{\ b} \mathcal{L}_{n}g_{\mu\nu},\ee
where ${\cal L}$ is the Lie derivative. 
The trace of the extrinsic curvature is given by 
\be K=h^{ab}K_{ab}=\nabla_\mu n^{\mu}.\ee
In terms of ADM variables, we have 
\be K_{ab}={1\over 2N}\left(\dot h_{ab}-\nabla_a N_b-\nabla_b N_a\right).\ee

We often use the decomposition
\be R=\ ^{(n-1)}R+\epsilon\left(K^2-K_{ab}K^{ab}\right)+2\epsilon\nabla_\alpha\left(n^\beta\nabla_\beta n^\alpha-n^\alpha K\right).\ee
Here $R$ is the Ricci scalar constructed from $g_{\mu\nu}$, and ${}^{(n-1)}R$ is the Ricci scalar constructed from the induced metric $h_{ab}$.  

\section{\label{foliationappendix} Foliation of spacetime}
\begin{figure}[h!]
\begin{center}
\includegraphics[height=6in]{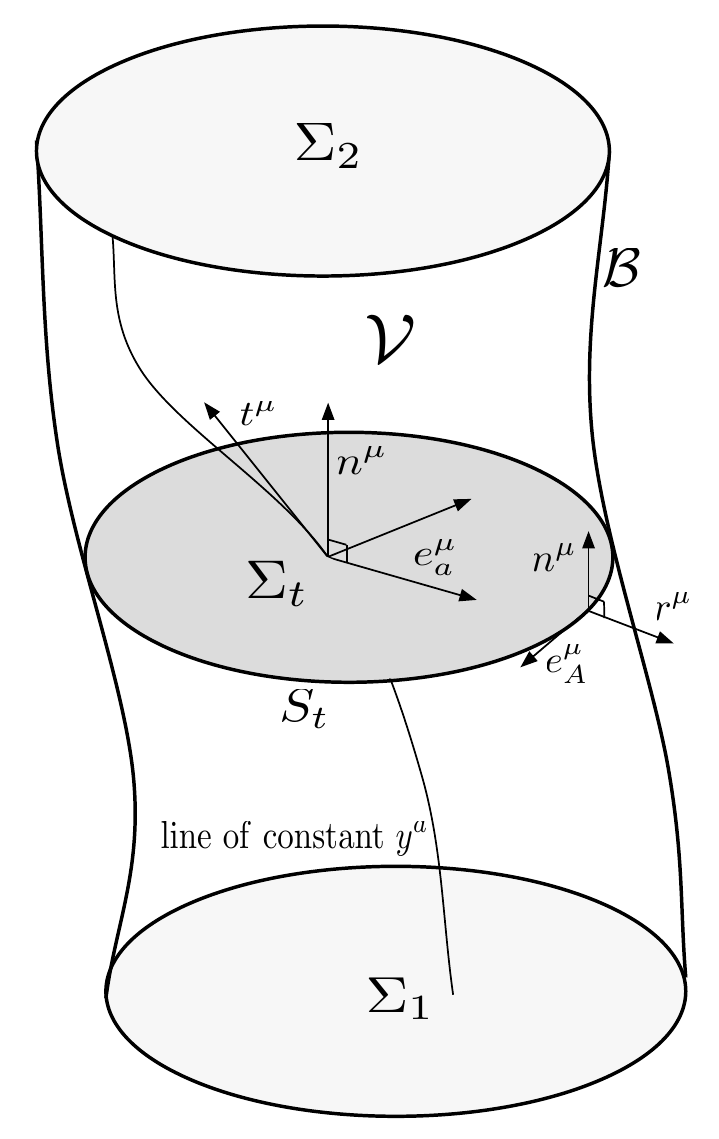}
\caption{Foliation of spacetime}
\label{leaf}
\end{center}
\end{figure}

Throughout this paper it is necessary to refer to spacetime, boundaries of spacetime, foliations of spacetime, and all of the geometrical quantities that these induce.  As such it is necessary to layout some standard terminology to describe the situation.  See figure \ref{leaf}.

Spacetime will be referred to as $\mathcal{V}$ bounded by $\partial\mathcal{V}$.  $\partial\mathcal{V}$ has three parts, a timelike boundary ${\cal B}$, a spacelike initial surface $\Sigma_1$ and a spacelike final surface $\Sigma_2$ .  We foliate the volume ${\cal V}$ with hypersurfaces $\Sigma_t$ by giving a global time function $t(x)$ and declaring the hypersurfaces to be its level sets.  The top and bottom surfaces $\Sigma_1$ and $\Sigma_2$ are to coincide with $\Sigma_{t_1}$ and $\Sigma_{t_2}$, for some times $t_1$ and $t_2$.  The normal vector to the $\Sigma_{t}$ is denoted $n^\mu$ and is future pointing.  The coordinates on $\Sigma_t$ are $y^a$.  The lapse and shift relative to this foliation are $N$, $N^a$.  The induced metric is $h_{ab}$ and the extrinsic curvature is $K_{ab}$.

The boundary ${\cal B}$ can be thought of as part of a foliation by timelike surfaces.  The coordinates on ${\cal B}$ are $z^i$.  The radially outward pointing normal vector is $r^\mu$.  We demand that the surfaces $\Sigma_t$ intersect ${\cal B}$ orthogonally, so that $g_{\mu\nu}r^\mu n^\nu=0$ on ${\cal B}$.  This implies that $r^\mu$ is parallel to $\Sigma_t$ on ${\cal B}$, so that $r^\mu=r^ae^\mu_{\ a}$ there, for some $r^a$.   The lapse and shift relative to this foliation are ${\cal N}$, ${\cal N}^i$.  The induced metric is $\gamma_{ij}$, and the extrinsic curvature is ${\cal K}_{ij}$.  

The intersections of $\Sigma_t$ with ${\cal B}$ are denoted $S_t=\Sigma_t\cap {\cal B}$.  These form a spacelike foliation of ${\cal B}$.  The coordinates on $S_t$ are $\theta^A$, the induced metric is $\sigma_{AB}$, and the extrinsic curvature of $S_t$ as embedded in $\Sigma_t$ is $k_{AB}$.  

\section{Conformal transformation formulae\label{conformalformulae}}
Here we collect some formulae on conformal transformations of the metric.  
\be \tilde g_{\mu\nu}=\omega^2(x)g_{\mu\nu},\ \ \ \omega(x)>0.\ee
\be \tilde g^{\mu\nu}=\omega^{-2}(x)g^{\mu\nu}.\ee
The connection transforms as 
\be \tilde\Gamma^\rho_{\mu\nu}=\Gamma^\rho_{\mu\nu}+C^\rho_{\ \mu\nu},\ee
\be C^\rho_{\ \mu\nu}=\omega^{-1}\left(\delta^\rho_\mu\partial_\nu\omega+\delta^\rho_\nu\partial_\mu\omega-g_{\mu\nu}g^{\rho\sigma}\partial_\sigma\omega\right).\ee
The curvature scalar transforms as 
\be \tilde R=\omega^{-2} R-2(n-1)g^{\alpha\beta}\omega^{-3}\nabla_\alpha\nabla_\beta\omega-(n-1)(n-4)g^{\alpha\beta}\omega^{-4}\partial_\alpha\omega\partial_\beta\omega.\ee
Some convenient covariant derivative transformations are 
\be \tilde\nabla_\mu\tilde\nabla_\nu\phi= \nabla_\mu\nabla_\nu\phi-\left(\delta^\alpha_\mu\delta^\beta_\nu+\delta^\beta_\mu\delta^\alpha_\nu-g_{\mu\nu}g^{\alpha\beta}\right)\omega^{-1}\partial_\alpha\omega\partial_\beta\phi.\ee
\be \tilde\square\phi=\omega^{-2}\square\phi+(n-2)g^{\alpha\beta}\omega^{-3}\partial_\alpha\omega\partial_\beta\phi.\ee

The inverse transformations are 
\be \label{rtildetransform} R=\omega^{2} \tilde  R+2(n-1)\tilde g^{\alpha\beta}\omega\tilde\nabla_\alpha\tilde\nabla_\beta\omega-n(n-1)\tilde g^{\alpha\beta}\partial_\alpha\omega\partial_\beta\omega.\ee

\be\nabla_\mu\nabla_\nu\phi= \tilde\nabla_\mu \tilde\nabla_\nu\phi+\left(\delta^\alpha_\mu\delta^\beta_\nu+\delta^\beta_\mu\delta^\alpha_\nu-\tilde g_{\mu\nu}\tilde g^{\alpha\beta}\right)\omega^{-1} \partial_\alpha\omega\partial_\beta\phi.\ee
\be\square\phi=\omega^{2} \tilde\square\phi-(n-2)\tilde g^{\alpha\beta}\omega\partial_\alpha\omega\partial_\beta\phi.\ee

The various 3+1 dimensional quantities transform as follows
\be \tilde h_{ab}=\omega^2 h_{ab},\ \ \ \tilde N^a=N^a,\ \ \ \tilde N=\omega N,\ee
\be \tilde h^{ab}=\omega^{-2} h^{ab},\ \ \ \tilde N_a=\omega^2 N_a,\ \ee
\be \tilde n^\mu=\omega^{-1}n^\mu,\ \ \ \tilde n_{\mu}=\omega n_{\mu},\ee
\be \tilde K_{ab}=\omega K_{ab}+h_{ab}n^\mu\partial_\mu\omega,\ee
\be \tilde K=\omega^{-1}K+\omega^{-2}(n-1)n^\mu\partial_\mu\omega.\ee

\section{\label{fundamentaltheorem} The fundamental theorem of auxiliary variables}

Here we recall some facts about auxiliary fields in classical field theory.  For more, see for example \cite{Henneaux:1991my,Henneaux:1992ig}.

Suppose the lagrangian depends on two sets of fields $\phi^i$ and $\chi^A$, $\mathcal{L}=\mathcal{L}([\phi],[\chi],x)$, (here $[\ ]$ stands for dependence on the fields and any of its spacetime derivatives of arbitrary but finite order) and that the equations of motion for $\chi^A$ can be solved in terms of the $\phi^i$, 
\be\frac{\delta^{EL}\mathcal{L}}{\delta\chi^A}=0\Rightarrow \chi^A=\chi^A([\phi],x).\ee
Plugging these relations back into $\mathcal{L}$, we get a lagrangian depending only on the $\phi^i$, which we call $\bar{\mathcal{L}}$,
\be \bar{\mathcal{L}}([\phi],x)\equiv \mathcal{L}([\phi],[\chi([\phi],x)],x).\ee

We will christen the following the fundamental theorem of auxiliary variables.
\begin{itemize}
\item The equations of motion derived from $\mathcal{L}$ and $\bar{\mathcal{L}}$ are equivalent in the $\phi^i$ sector, i.e. if $\phi^i(x)$ is a solution to $\frac{\delta^{EL}{\mathcal{L}}}{\delta\phi^i}=0$, then it is also a solution to $\frac{\delta^{EL}\bar{\mathcal{L}}}{\delta\phi^i}=0$, and vice versa, if $\phi^i(x)$ is a solution to $\frac{\delta^{EL}\bar{\mathcal{L}}}{\delta\phi^i}=0$, then it is also a solution to $\frac{\delta^{EL}{\mathcal{L}}}{\delta\phi^i}=0$, with the extension to the $\chi^A$ given by the $\chi^A([\phi],x)$ obtained from solving the $\chi^A$ equations of motion.
\end{itemize}
Proving this is a matter of convincing yourself that you can extremize a function of several variables either by extremizing with respect to all the variables, or by first extremizing with respect to a few, and then with respect to the rest.  

The other important property that hold when eliminating auxiliary variables is that both the global and gauge symmetry groups of ${\cal L}$ and $\bar{\cal L}$ are the same.  One does not lose or gain symmetries by eliminating or adding auxiliary variables.  

Notice that in the fundamental theorem, there is no requirement that the auxiliary fields be solved for algebraically.  However, when this is not the case, one must be careful about boundary contributions.  To illustrate this, we reprint here a nice example taken from the exercises of chapter one of \cite{Henneaux:1992ig}.

Consider the action for two variables $q(t)$ and $A(t)$ with values fixed on the endpoints, 
\be S=\int _{t_1}^{t_2}dt\ \half\left(\dot q^2+{\dot A^2\over q^2}\right).\ee
The equations of motion for $q$ and $A$ are, respectively,
\be \ddot q+{\dot A^2\over  q^3}=0,\ \ \ {d\over dt}\left(\dot A\over q^2\right)=0.\ee
Notice that the $A$ equation implies $\dot A=c q^2$, where $c$ is a constant of integration.  Plugging this back into the action, we obtain 
\be S=\int _{t_1}^{t_2}dt \ \half\left(\dot q^2+c^2 q^2\right), \ee
which gives naively for the $q$ equation of motion $\ddot q-c^2 q=0$.  Looking back at the original $q$ equation of motion, and plugging in $\dot A=c q^2$, we obtain $\ddot q+c^2 q=0$, a contradiction for $c^2\not=0$.   

It is often said that auxiliary variables can only be eliminated if their equations of motion can be solved algebraically, and ``counterexamples'' like this are quoted to illustrate this.  However, if we are careful to take into account the endpoints, we can resolve the problem.  Consider again the equation of motion for $A$, ${d\over dt}\left(\dot A\over q^2\right)=0$.  We must solve this subject to the endpoint conditions $A(t_1)=A_1$, $A(t_2)=A_2$.  Integrating both sides of the solution $\dot A=c q^2$, we find that the constant $c$ actually depends non-locally on $q$, as well as on the endpoint data for $A$, 
\be  c={A_2-A_1\over \int _{t_1}^{t_2} q(t)^2dt}.\ee

Plugging this more careful result into the action, we obtain a non-local action for $q$, 
\be S=\int _{t_1}^{t_2}dt\ \half\left(\dot q^2+\left({A_2-A_1\over \int _{t_1}^{t_2} q(t')^2dt'}\right)^2 q^2\right).\ee
Varying this carefully with respect to $q$, we recover the correct (non-linear, non-local) equation of motion $\ddot q+c^2 q=0$.

\providecommand{\href}[2]{#2}\begingroup\raggedright\endgroup

\end{document}